\documentclass[final,3p,times,twocolumn]{elsarticle}
\usepackage{amssymb}

\newcommand{\lat}{\emph{Fermi}-LAT}
\newcommand{\chandra}{\emph{Chandra}}
\newcommand{\xmm}{XMM-\emph{Newton}}
\newcommand{\suzaku}{\emph{Suzaku}}
\newcommand{\integral}{\emph{INTEGRAL}}
\newcommand{\asca}{\emph{ASCA}}
\newcommand{\hess}{\emph{H.E.S.S.}}
\newcommand{\magic}{\emph{MAGIC}}
\newcommand{\veritas}{\emph{VERITAS}}
\newcommand{\astrosat}{\emph{ASTROSAT}}
\newcommand{\nustar}{\emph{NuSTAR}}
\newcommand{\erosita}{\emph{e-ROSITA}}
\newcommand{\astroh}{\emph{ASTRO-H}}
\newcommand{\gems}{\emph{GEMS}}
\newcommand{\athena}{\emph{Athena}}
\newcommand{\loft}{\emph{LOFT}}
\newcommand{\ibis}{IBIS/\emph{ISGRI}}
\newcommand{\rosat}{\emph{ROSAT}}
\newcommand{\hegra}{\emph{HEGRA}}
\newcommand{\agile}{\emph{AGILE}}
\newcommand{\swift}{\emph{SWIFT}}
\newcommand{\cgro}{\emph{CGRO}}
\newcommand{\cta}{\emph{CTA}}

\begin{document}

\begin{frontmatter}


\title{Multiwavelength Astronomy and CTA: X-rays}
\author[isas]{Tadayuki Takahashi}
\author[slac]{Yasunobu Uchiyama}
\author[isas,oauj]{{\L}ukasz Stawarz}
\address[isas]{Institute of Space and Astronautical Science, JAXA, 3-1-1 Yoshinodai, Chuo-ku, Sagamihara, Kanagawa 252-5210, Japan}
\address[slac]{SLAC National Accelerator Laboratory, 2575 Sand Hill Road, Menlo Park CA 94025, USA}
\address[oauj]{Astronomical Observatory, Jagiellonian University, ul. Orla 171, 30-244 Krak\'ow, Poland}

\begin{abstract}
We discuss how future X-ray instruments which are under development can contribute to our understanding of the non-thermal Universe. Much progress has been made in the field of X-ray Astronomy recently, thanks to the operation of modern X-ray telescopes such as \chandra, \xmm, \suzaku, and \swift, but more in-depth investigation awaits future missions. These future missions include \astrosat, \nustar, \erosita, \astroh\ and \gems, which will be realized in the next decade, and also much larger projects such as \athena\ and \loft, which have been proposed for the 2020's. All of those are expected to bring a variety of novel observational results regarding astrophysical sources of high-energy particles and radiation, i.e. supernova remnants, neutron stars, stellar-mass black holes, active galaxies, and clusters of galaxies among others. The operation of the future X-ray instruments will proceed in parallel with the operation of \lat\ and the Cherenkov Telescope Array. We emphasize that the synergy between the X-ray and $\gamma$-ray observations is particularly important, and that the planned X-ray missions, when in conjunction with the modern $\gamma$-ray observatories, will indeed provide a qualitatively better insight into the high-energy Universe.
\end{abstract}


\end{frontmatter}

\section{Introduction}
\label{Sec:Intro}

X-ray observations using space telescopes have revealed that the Universe is full of high-temperature phenomena reaching 10 to 100 million degrees, which nobody had imagined before the advent of X-ray Astronomy. The X-ray band is capable of probing extreme conditions of the Universe such as the proximity of black holes or the surface of neutron stars, as well as observing exclusively the emission from high-temperature gas and selectively the emission from accelerated electrons. In recent years, \chandra, \xmm, \suzaku\ and other X-ray missions have made great advances in X-ray Astronomy. We have obtained knowledge which revolutionized our understanding of the high energy Universe and learned that phenomena observed in the X-ray band are deeply connected to those observed in other wavelengths from radio to $\gamma$-rays.

X-rays of synchrotron origin are of special interest because they are generally produced in extreme cosmic accelerators, which can accelerate particles up to and above $10^{12}$\,eV energies. X-rays carry information not only about the directly accelerated electrons, but also about hadrons, through the synchrotron radiation of secondary $e^{\pm}$ pairs produced at interactions of accelerated protons and nuclei with ambient gas and radiation fields. The \asca\ X-ray satellite and the first generation TeV telescopes demonstrated the close relationship between X-rays and TeV $\gamma$-rays from objects such as ``high frequency peaked'' blazars and young supernova remnants \cite{Ref:Mrk421_TT,Ref:RXJ1713_HESS}. Since then, tremendous achievements which show the link between these two frequency bands have been done with the present X-ray satellites and the second generation TeV telescopes including \hess, \magic, and \veritas. Recently, hard X-ray observations are becoming more important, since this is the energy band where non-thermal emission could overwhelm thermal X-ray emission in sources like galaxy clusters and supernova remnants. The hard X-ray emission, if detected, traces the sites of particle acceleration in such objects, and gives important information about the particle acceleration mechanisms involved. As discussed below, the hard X-ray telescopes onboard \astroh\ and \nustar, with sensitivity two orders of magnitude better than the present missions, are capable of solving various scientific questions to understand the non-thermal Universe.

In order to study turbulence, magnetic fields, and relativistic particles in various astrophysical systems, and to draw a more complete picture of the high energy Universe, observations by a spectrometer with an extremely high resolution capable of measuring the bulk plasma velocities and/or turbulence with a resolution corresponding to a speed of $\sim 100$\,km\,s$^{-1}$ are desirable. In galaxy clusters, X-ray hot gas is trapped in a gravitational potential well and shocks and/or turbulence are produced as smaller substructures with their own hot gas halos fall into and merge with the dominant cluster. Large scale shocks can also be produced as gas from the intracluster medium falls into the gravitational potential of a cluster. The bulk motions and turbulences are in turn responsible for acceleration of particles to very high energies, which is manifested via non-thermal emission processes, best studied with sensitive hard X-ray and $\gamma$-ray measurements.

Understanding the non-thermal phenomena in the Universe is one of the key goals of modern astrophysics. The origin of galactic and extragalactic cosmic rays and their roles in the history of the Universe still remain unsolved. In this paper, we will discuss contributions by future X-ray missions which are under development in conjunction with possible synergy with the next-generation TeV $\gamma$-ray observatory, the Cherenkov Telescope Array (\cta).
 
\section{Future X-ray Missions}
\label{Sec:Missions}

A number of new X-ray missions which are expected to revolutionize the current understanding of the high energy Universe are being developed and planned. In the next decade, \astrosat\ \cite{Ref:ASTROSAT}, \nustar\ \cite{Ref:NuSTAR}, \erosita\ \cite{Ref:eROSITA}, \astroh\ \cite{Ref:ASTRO-H} and \gems\ \cite{Ref:GEMS} will be realized. Among them, the 6th Japanese X-ray satellite \astroh, to be launched in 2014, is the next major international X-ray mission which will be operated as an observatory. Much larger missions, such as \athena\ \cite{Ref:Athena} and \loft\ \cite{Ref:LOFT}, have been proposed for the 2020's.

%%%%%%%%%%%%%%%%%%%%%%%%%%%%%%%%%%%
\begin{figure}[t]
\centerline{\includegraphics[scale=0.32,angle=90]{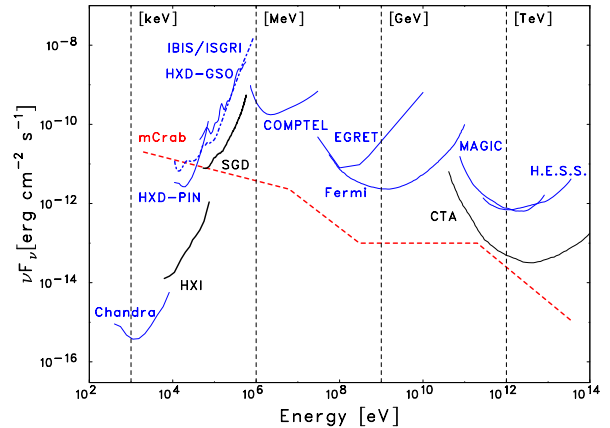}}
\caption{Differential sensitivities of different X-ray and $\gamma$-ray instruments for an isolated point source. 
Lines for the \chandra/ACIS-S, the \suzaku/HXD (PIN and GSO), the \integral/IBIS (from the 2009 IBIS Observer's Manual), and the \astroh/HXI,SGD are the $3\sigma$ sensitivity curves for 100 ks exposures. A spectral bin with $\Delta E/E = 1$ is assumed for \chandra\ and $\Delta E/E = 0.5$ for the other instruments. 
Note that the \xmm\ instruments have a slightly better sensitivity than \chandra\ for 100 ks, while \swift/BAT is characterized by almost the same sensitivity limit as \ibis\ within the range from 15\,keV up to $\sim 300$\,keV.
The sensitivities of the COMPTEL and EGRET instruments correspond to the all-lifetime all-sky survey of \cgro. The curve denoting \lat\ is the pre-launch sensitivity evaluated for the $5 \sigma$ detection limit at high Galactic latitudes with 1/4-decade ranges of energy in a one-year dataset \cite{Atwood2009}. The curves depicting the \magic\ Stereo system \cite{Carmona2011} and \hess\ are given for $5\sigma$ detection with $>10$ excess photons after 50\,h exposure. The simulated \cta\ configuration C sensitivity curve for 50\,h exposure at a zenith angle of $20$\,deg is taken from \cite{CTA}. Red dashed line denotes the differential energy flux corresponding to the mCrab unit in various energy ranges as adopted in the literature.}
\label{Fig:Sensitivity}
\end{figure}
%%%%%%%%%%%%%%%%%%%%%%%%%%%%%%%%%%%

\astrosat\ is a multi-wavelength astronomy mission carrying four X-ray instruments, which will be placed in a 650-km, near-equatorial orbit. It will provide data mainly in the area of X-ray timing and broadband spectroscopy covering the energy range $0.3-150$\,keV, with emphasis on hard X-rays. Diffuse UV studies can also be carried out with an onboard UV telescope.

\nustar\ and \astroh\ will carry the first focusing hard X-ray telescopes with graded multilayer reflecting surfaces that operate in an energy range of  $5-80$\, keV. 
Imaging  and especially focusing instruments have two tremendous advantages. Firstly, the volume of the focal plane detector can be made much smaller than for non-focusing instruments, so reducing the absolute background level since the background flux generally scales with the size of the detector. Secondly, the residual background, often time-variable, can be measured simultaneously with the source, and can be reliably subtracted.  

%%%%%%%%%%%%%%%%%%%%%%%%%%%%%%%%%%%
\begin{figure}
\centerline{\includegraphics[scale=0.27,clip,angle=270]{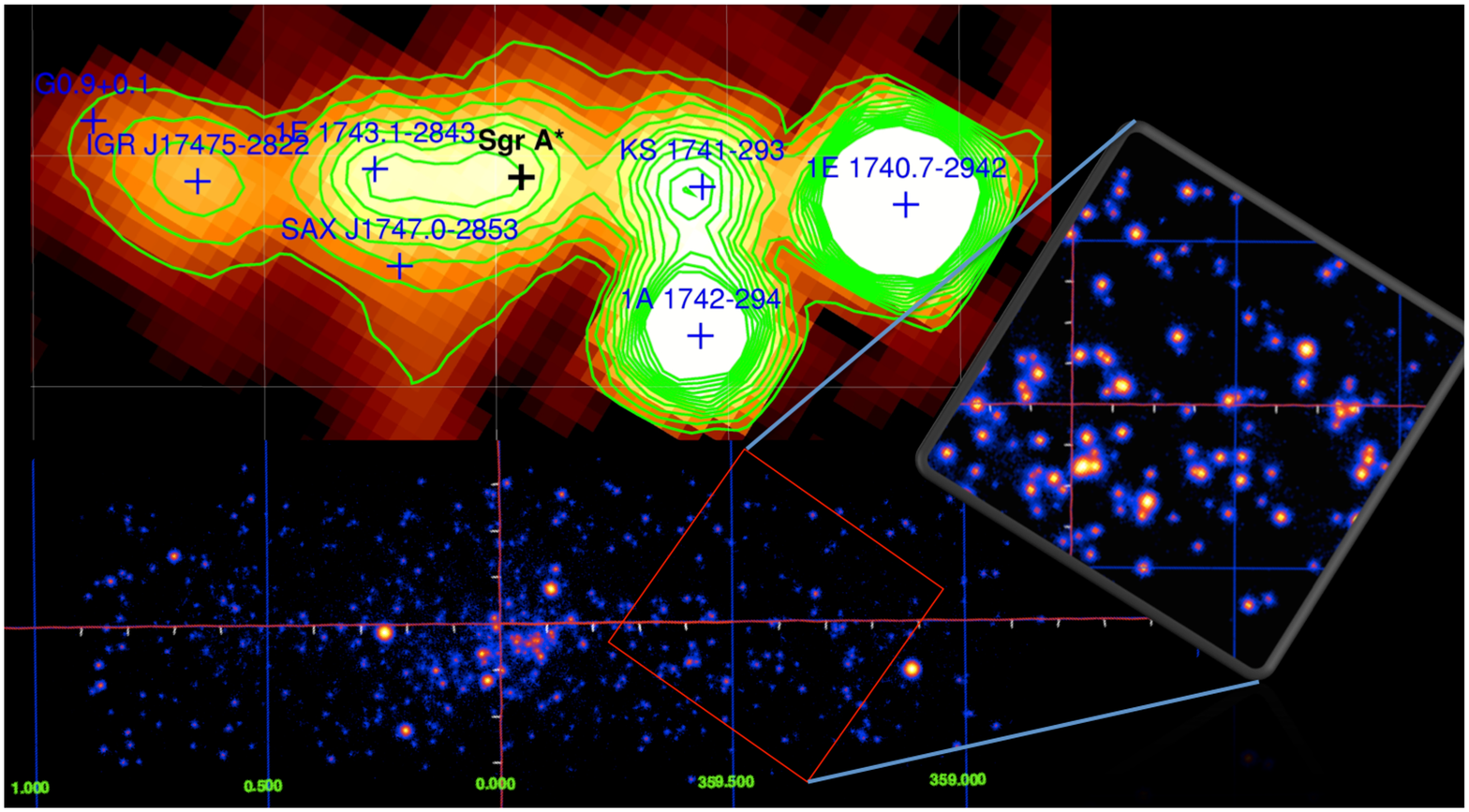}}
\caption{The \ibis\ significance mosaic of the Galactic Center region in the $20-40$\,keV energy range \cite{Ref:INTEGRAL_GC} is shown in the top of the image. In the bottom of the image the expected performance of hard X-ray observations of the same region with \nustar\ is presented. The simulation does not include molecular clouds but focuses on the population of X-ray binaries, with a depth of 12ksec/pixel or 6\,$\mu$Crab (courtesy of F.A. Harrison).}
\label{Fig:NuSTAR}
\end{figure}
%%%%%%%%%%%%%%%%%%%%%%%%%%%%%%%%%%%

As shown in Figure\,\ref{Fig:Sensitivity}, the sensitivity to be achieved by \astroh\ (and similarly \nustar) is about two orders of magnitude improved compared to previous collimated or coded mask instruments that have operated in this energy band (Figure\,\ref{Fig:NuSTAR}). This will bring a breakthrough in our understanding of hard X-ray spectra of celestial sources in general. With this sensitivity, $30-50$\,\% of the hard X-ray Cosmic Background would be resolved. This will enable us to track the evolution of active galaxies with accretion flows which are heavily obscured, in order to accurately assess their contribution to the Cosmic X-ray Background over cosmic time. In addition, simultaneous observations of blazar-type active galaxies with \lat\ and the TeV $\gamma$-ray telescopes are of vital importance to study particle acceleration in relativistic jets (see \S\,\ref{Sec:AGN}).  

\begin{figure}
\centerline{\includegraphics[scale=0.5,clip]{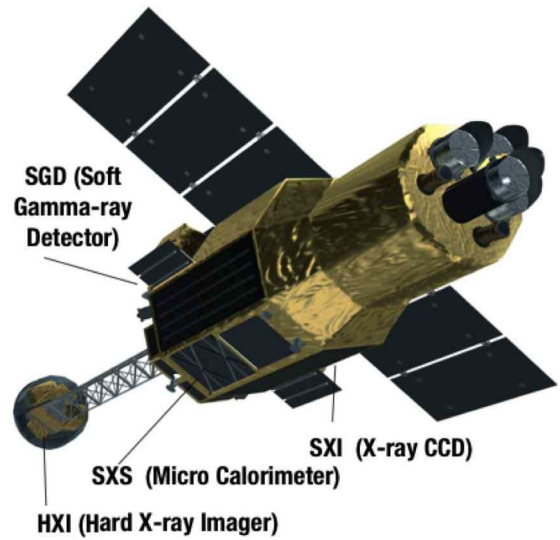}}
\caption{Schematic view of the \astroh\ satellite. The total mass at launch will be $\sim 2700$\,kg. \astroh\ will be launched into a circular orbit with altitude $500-600$\,km, and inclination $\sim 31$\,degrees \cite{Ref:ASTRO-H}.}
\label{Fig:ASTRO-H}
\end{figure}

In addition to the hard X-ray telescopes, \astroh\ will carry two Soft X-ray Telescopes, one with a micro-calorimeter spectrometer array with excellent energy resolution (Soft X-ray Spectrometer; SXS), and the other with a large area CCD in their respective focal planes (Figure\,\ref{Fig:ASTRO-H}). The spectroscopic capability of X-ray micro-calorimeters is unique in X-ray astronomy, since no other spectrometers can achieve high energy resolution, high quantum efficiency, and spectroscopy for spatially extended sources at the same time. Imaging spectroscopy with an energy resolution $< 7$\,eV by the SXS of extended sources can reveal line broadening and Doppler shifts due to turbulent or bulk velocities of the X-ray emitting plasma. This capability enables the determination of the level of turbulent pressure support in clusters, supernova ejecta dispersal patterns, the structure of active galactic and starburst winds, and the spatially dependent abundance pattern in clusters and elliptical galaxies. The SXS can also measure the optical depths of resonance absorption lines, from which the degree and spatial extent of turbulence can be inferred. 

%%%%%%%%%%%%%%%%%%%%%%%%%%%%%%%%%%%
%%%%%%%%%%%%%%%%%%%%%%%%%%%%%%%%%%%

\begin{figure}
\centerline{\includegraphics[scale=0.575,clip]{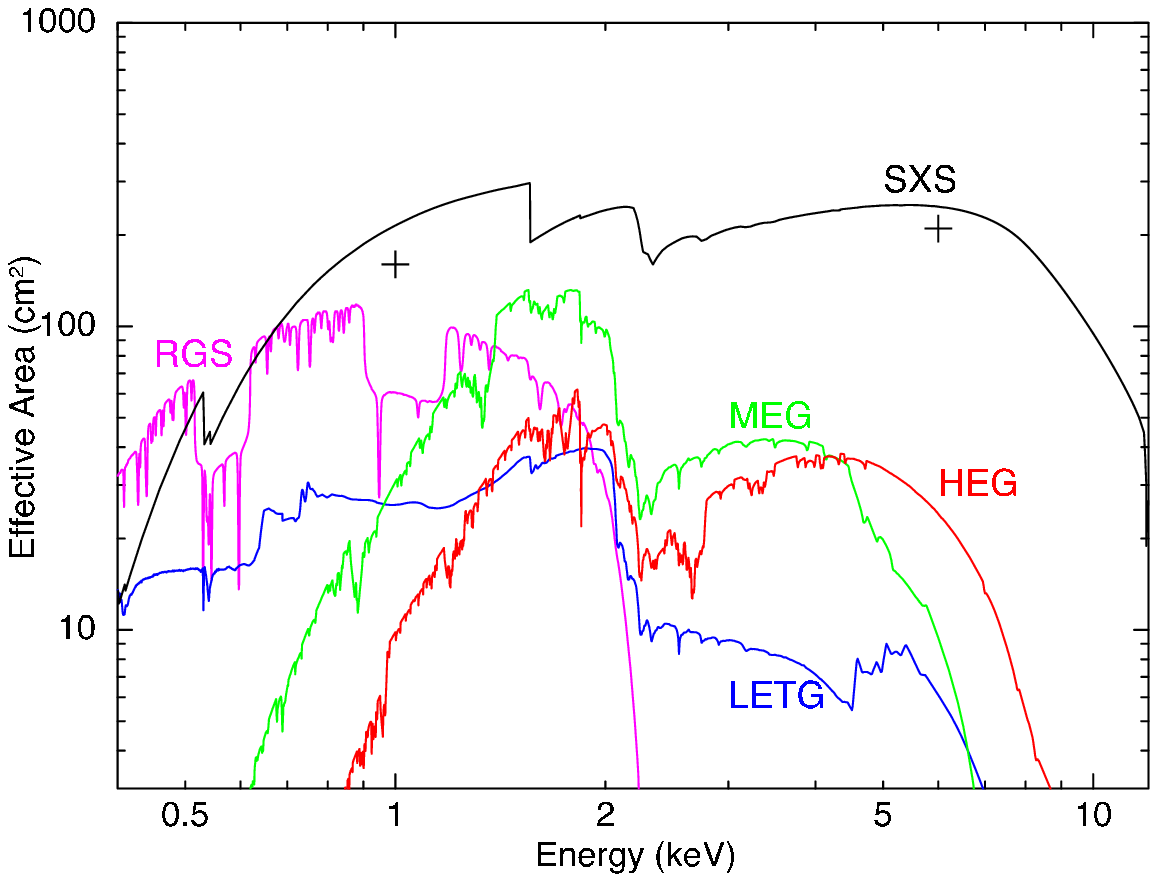}}
\centerline{\includegraphics[scale=0.575,clip]{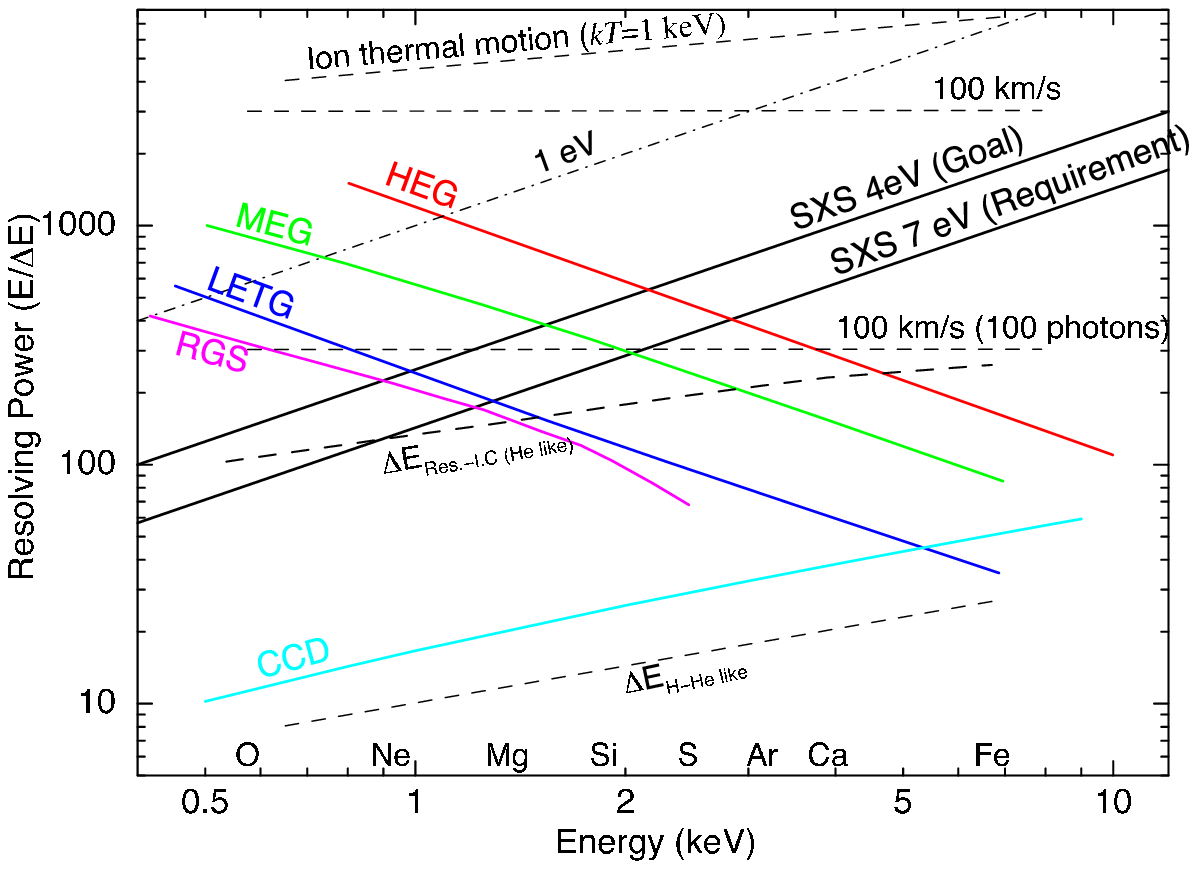}}
\caption{{\bf (a)} Effective areas of high-resolution X-ray spectroscopy missions as functions of X-ray energy. The curve for the \astroh\ SXS is the present best estimate for a point source. The two crosses show the mission requirements. The \xmm\ RGS effective area is a sum of first order of the two instruments (RGS-1 and RGS-2). The effective areas of LETG, MEG and HEG onboard \chandra\ are sums of first order dispersions in $\pm$ directions. {\bf (b)} Resolving power of the \astroh\ SXS as a function of X-ray energy for the two cases, 4\,eV resolution (goal) and 7\,eV (requirement). The resolving power of high resolution instruments onboard \chandra\ and \xmm\ and typical resolving power of X-ray CCD cameras are also shown for comparison \cite{Ref:SXS}. }
\label{Fig:XRS}
\end{figure}

In combination with a high throughput X-ray telescope, the SXS improves on the \chandra\ and \xmm\ grating spectrometers in two important ways. At $E> 2$\,keV, the SXS is both more sensitive and has higher resolution (Figure\,\ref{Fig:XRS}), especially in the Fe K band where the SXS has 10 times the collecting area and much better energy resolution, giving a net improvement in sensitivity by a factor of 30 over \chandra. In addition the SXS uniquely performs high-resolution spectroscopy of extended sources. In contrast to a grating, the spectral resolution of the calorimeter is unaffected by source's angular size because it is non-dispersive. 

In order to extend the energy coverage to the soft $\gamma$-ray region up to 600\,keV, the Soft Gamma-ray Detector (SGD) will be implemented as a non-focusing detector onboard \astroh. The SGD measures soft $\gamma$-rays via reconstruction of the Compton scattering in the Compton camera, covering  an energy range of $40-600$\,keV with sensitivity at $300$\,keV of more than 10 times better than the \suzaku\ Hard X-ray Detector. The SGD is capable of measuring the polarization of celestial sources brighter than a few times 1/100 of the Crab Nebula and polarized above $\sim 10\,\%$. This capability is expected to yield polarization measurements in several celestial objects, providing new insights into properties of soft $\gamma$-ray emission processes.

The Gravity and Extreme Magnetism Small Explorer (\gems) is an astrophysical observatory dedicated to X-ray polarimetry ($2-10$\,keV) and is being developed for launch in 2014. \gems\ will perform the first sensitive X-ray polarization survey of several classes of X-ray emitting sources characterized by strong gravitational or magnetic fields. It has been recognized for a long time that X-ray polarization measurements can provide unique diagnosis of the strong fields near compact objects. The prime scientific objectives of \gems\ are to determine the effects of the spin of black holes, the configurations of the magnetic fields of magnetars, and the structure of the supernova shocks which accelerate cosmic rays. In the cases of both stellar black holes and supermassive black holes, sensitivity to $1\,\%$ polarization is needed to make diagnostic measurements of the net polarizations predicted for probable accretion disk and corona models. \gems\ can reach this goal for several Seyfert galaxies and quasars and measure the polarizations of representatives of a variety of other classes of X-ray sources, such as rotation-powered and accretion-powered pulsars. 

%%%%%%%%%%%%%%%%%%%%%%%%%%%%%%%%%%%
%%%%%%%%%%%%%%%%%%%%%%%%%%%%%%%%%%%

\erosita\ will be the primary instrument onboard the Russian ``Spectrum-Roentgen-Gamma'' (SRG) satellite which will be launched in 2013 and placed in an L2 orbit \cite{Ref:eROSITA}. The \erosita\ mission will perform the first imaging all-sky survey in the medium energy X-ray range up to 10\,keV with an unprecedented spectral and angular resolution. The \erosita\ sensitivity during the all-sky survey for four years will be approximately 30 times \rosat. In the all-sky survey, the typical flux limit will be $\sim 10 ^{-14}$\,erg\,cm$^{-2}$\,s$^{-1}$ and $\sim 3 \times 10^{-13}$\,erg\,cm$^{-2}$\,s$^{-1}$ in the $0.5-2$\,keV and $2-10$\,keV energy bands, respectively. At these fluxes the X-ray sky is dominated by active galaxies and clusters, which can be separated with an angular resolution of $25-30$\,arcsec. The proposed survey will identify 50,000--100,000 clusters depending on the capabilities in disentangling moderately-low extended sources from active galaxies. Concerning the number of active galaxies, the $\log N - \log S$ measurement in moderately wide field surveys, like XMM-COSMOS, can be used to predict detections of $(3-10) \times 10^{6}$ sources, up to $z \sim 7-8$, depending on the detection threshold.

\section{Supernova Remnants}
\label{Sec:SNRs}

\subsection{X-ray Study of Supernova Remnants}

X-ray imaging and spectroscopic observations play an important role in understanding TeV $\gamma$-ray emission from supernova remnants (SNRs). Generally, SNRs are studied with X-ray instruments in the following contexts: (i) SNRs are believed to be the primary sources of galactic cosmic rays (CRs) up to the knee (at $\sim 3\times 10^{15}$\,eV in the all-particle CR spectrum) and possibly beyond it; (ii) SNRs are the best sites to study particle acceleration (particularly, diffusive shock acceleration, DSA) processes, which should have wide applications in astrophysics; (iii) supernovae are important sources of chemical elements in the Universe; X-ray line spectroscopy of SNRs can probe nucleosynthesis taking place in the interior of stars and during supernova explosions, being complementary to a late phase optical spectroscopy of supernovae; (iv) SNRs are major sources of kinetic energy and turbulence of interstellar gas, thereby affecting the gas structures of our Galaxy and also star formation. The synergies between X-ray and TeV observations of SNRs primarily lie in the areas of particle acceleration and the origin of Galactic CRs. 

\begin{figure}
\begin{center}
\includegraphics*[scale=0.4]{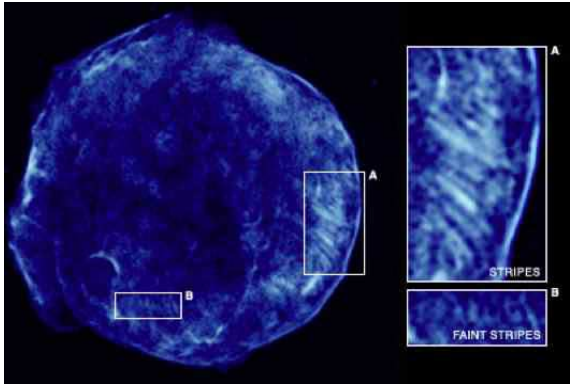}
\end{center}
\caption{Deep \chandra\ $4-6$\,keV image of Tycho's SNR \cite{Eriksen11}.
Bright features are due to synchrotron radiation produced by multi-TeV electrons.}
\label{fig:Eriksen}
\end{figure}

In young SNRs, both nonthermal and thermal X-ray emissions can be observed. Synchrotron radiation by TeV electrons accelerated at strong shock fronts, first identified with the \asca\ satellite \cite{Koyama95}, is currently the only established channel of nonthermal X-radiation in SNRs. Observations of synchrotron-emitting X-ray filaments provide key information about particle acceleration and magnetic field amplification processes (see \S\,\ref{sec:MFA} and \S\,\ref{sec:Bykov}). For instance, a recent deep \chandra\ map of Tycho's SNR has revealed an interesting spatial feature, ``stripes'', of synchrotron X-ray emission (see Figure\,\ref{fig:Eriksen}) which may be taken as signatures of magnetic field amplification and associated acceleration of CR protons and nuclei up to $\sim 10^{15}$\,eV \cite{Eriksen11}.

%%%%%%%%%%%%%%%%%%%%%%%%%%%%%%%%%%%
%%%%%%%%%%%%%%%%%%%%%%%%%%%%%%%%%%%

\begin{figure}
\begin{center}
\includegraphics*[scale=0.4]{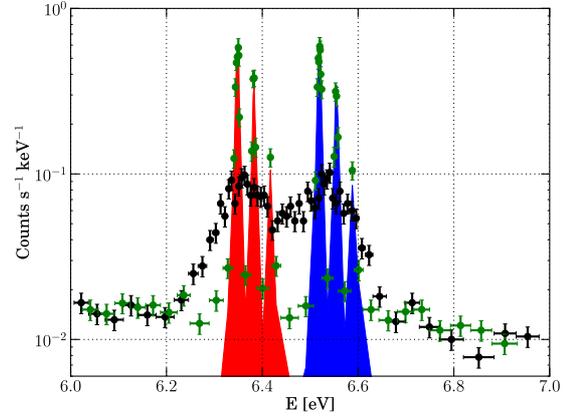}
\end{center}
\caption{
Simulated \astroh\ SXS spectrum (black points) around the iron K-shell complex of Tycho's SNR for an exposure of 100\,ks 
(taken from \astroh\ Quick Reference \texttt{http://\-astro-h.isas.jaxa.jp/\-researchers/\-news/\-2010/\-1119\_e.html}). 
For a reference, a simulated spectrum with no 
thermal Doppler broadening is also shown (green points). 
Iron K lines from a blob that is receding are red-shaded, while 
lines from an approaching blob are blue-shaded.  }
\label{fig:Tycho}
\end{figure}

Thermal components include the line and continuum (bremsstrahlung) emissions from shock-heated interstellar/circumstellar medium and from the hot ejecta heated by reverse shocks. Dissipation at a shock occurs through wave-particle interactions %%(namely, a collisionless shock) 
and shock-heating is inevitably connected with shock-acceleration. As discussed below, X-ray diagnostics of shocked plasma in SNRs aids in understanding of TeV $\gamma$-ray emission. Line emissions come from electron-collisional excited ions, dominated by alpha elements like O, Ne, Mg, Si, and S, and the iron peak elements like Fe and Ni. 
Shocked plasmas in young SNRs generally do not 
reach collisional ionization equilibrium (CIE) and they are under-ionized. 
It takes $\tau \equiv nt \sim 10^{12}$\,s\,cm$^{-3}$ to reach CIE.
Collisional heating of electrons by ions is also a slow process and consequently 
electron-ion temperature equilibrium is not reached. 
The degree of electron-proton temperature equilibration at the shock front 
is determined by collisionless heating via collective plasma processes, which 
are not well understood. The analysis of Balmer-dominated 
optical spectra of partially ionized shocks indicates a temperature ratio of 
$T_e/T_p = 0.05 - 0.1$ for
a  high shock speed ($v > 1000$\,km\,s$^{-1}$) \cite{Ref:Balmer-dominated}.
On the other hand, ion temperatures at fully ionized shocks (dominant for young SNRs) 
have never been well determined.
The excellent resolution of the \astroh\ SXS offers an opportunity to measure a temperature of shocked irons in young SNRs, 
a real breakthrough in understanding the physics of shock heating. 
Figure~\ref{fig:Tycho} presents simulated spectra of SXS observations of the central portion of Tycho's SNR. The spectra (black points) assume 
two plasma blobs that are receding and approaching to us with $\pm 4000\ \rm km\ s^{-1}$ \cite{Ref:Hayato} with the same 
parameters, an iron temperature of $kT_{\rm Fe} = 3$ MeV (mass-proportional heating),  
an electron temperature of $kT_e = 5$ keV,  and an ionization parameter 
of $\tau = 0.9\times 10^{10}$\,s\,cm$^{-3}$. For a reference, a simulated spectrum with no thermal Doppler broadening is also shown (green points).

%%%%%%%%%%%%%%%%%%%%%%%%%%%%%%%%%%%
%%%%%%%%%%%%%%%%%%%%%%%%%%%%%%%%%%%

Only thermal X-ray emission has been observed in evolved SNRs (say, $>10^4$\,yr), usually from low temperature ($kT_e \sim 0.1$\,keV) CIE plasmas. Recently, X-ray emission from overionized (recombining) plasma \cite{Kawasaki02} has been observed in Mixed Morphology SNRs (MMSNRs) with the \suzaku\ satellite \cite{Yamaguchi09}. 
MMSNRs are mostly strong GeV $\gamma$-ray emitters \cite{Uchi11}. X-ray observations can be used to infer the dynamical evolution of such SNRs. Conducting sensitive searches for nonthermal bremsstrahlung in the hard X-ray band by \nustar\ and \astroh\ HXI will complement $\gamma$-ray measurements. 

\subsection{Synergy between X-ray and Gamma-ray Observations of SNRs} 

A supernova origin of CRs has long been a matter of active research since it was advocated by Baade and Zwicky in the 1930's \cite{BaadeZwicky34}. The current sophisticated paradigm is that diffusive shock acceleration (DSA) at collisionless shock waves of SNRs is responsible for the production of Galactic CRs up to the knee energy or even beyond \cite{MD01}, transferring $\sim 10\,\%$ of the explosion kinetic energy into the form of CR energy \cite{Hillas_Review05}. DSA is widely regarded as the standard mechanism for producing relativistic particles at collisionless shocks in various astrophysical objects.

$\gamma$-ray observations of SNRs provide the most straightforward way of addressing the SNR paradigm for the origin of CRs through the measurement of $\pi^0$-decay $\gamma$-rays \cite{DAV94}. \hess\ observations of TeV $\gamma$-ray emission from SNR RX\,J1713.7$-$3946 \cite{HESS_1713,HESS_1713_2} have revealed a TeV $\gamma$-ray morphology closely matching the synchrotron X-ray map (see Figure\,\ref{fig:RXJ1713}), providing a firm example of TeV $\gamma$-ray emission from an SNR shell. SNRs constitute one of the most populated classes of TeV sources in our Galaxy \cite{2009_TeVREview}. Given that the galactic CRs are energetically dominated by protons and that DSA models usually presume a very high efficiency of proton acceleration, finding evidence for the $\pi^0$-decay $\gamma$-rays is indispensable, but it has remained tantalizingly difficult mainly because radiation processes involving relativistic electrons (the so-called \emph{leptonic} components) could explain the $\gamma$-ray emission as well. 

{X-ray observations play an important role in constraining the origin of TeV $\gamma$-ray emission from shell-type SNRs. 
A synchrotron X-ray spectrum is tightly coupled to the IC $\gamma$-ray spectrum that is produced by the same population of accelerated electrons. Moreover, X-ray measurements provide information about the hydrodynamic structure (e.g., shock speed, gas density) and the magnetic field strength in a remnant, which are crucial to disentangle $\gamma$-ray emission mechanisms. 

%Recently observations with the \lat\ have started to provide an effective means to disentangle the leptonic (IC scattering and relativistic bremsstrahlung) and hadronic ($\pi^0$-decay) $\gamma$-ray components. For example, the GeV spectral shape of SNR Cassiopeia\,A can be better fitted by a $\pi^0$-decay spectrum \cite{CasA}, indicating the hadronic origin of the TeV $\gamma$-ray emission. 

\subsection{Magnetic Field Amplification in Young SNRs} 
\label{sec:MFA}

High angular resolution observations of young SNRs with \chandra\ suggest that strong shocks may be able to amplify the interstellar magnetic field by large factors. The narrow widths of synchrotron X-ray filaments \cite{Bamba05} could be due to rapid synchrotron cooling in the postshock flow \cite{VL03}. The magnetic field strength inferred from the X-ray filaments is typically $\sim 0.1$\,mG \cite{Voelk05}. An alternative explanation for the narrowness of the filaments is a fast magnetic field damping behind a shock \cite{Pohl05}. This scenario also requires similarly strong magnetic fields. Evidence for the amplified magnetic field comes also from the year-scale time variability of synchrotron X-ray filaments (Figure\,\ref{fig:RXJ1713}) \cite{Uchi07}. If the variability timescale represents the synchrotron cooling time, the magnetic field strength can be estimated to be as large as $\sim 1$\,mG. On the other hand, if the variability is due to intermittent turbulent magnetic fields \cite{Bykov09} (see \S\,\ref{sec:Bykov}), the time variability can be reconciled with a weaker magnetic field ($\sim 0.1$\,mG). 

%%%%%%%%%%%%%%%%%%%%%%%%%%%%%%%%%%%
\begin{figure}
\begin{center}
\includegraphics*[scale=0.3]{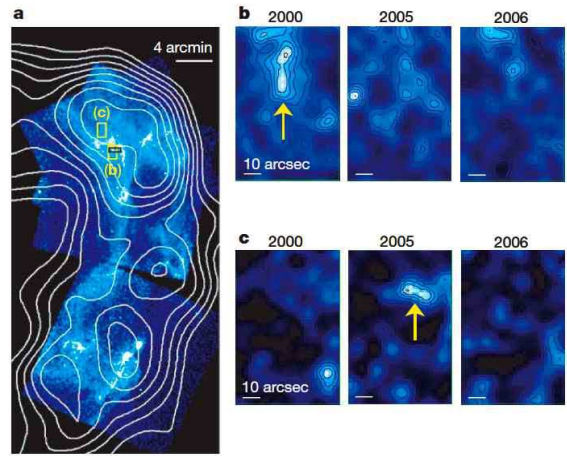}
\end{center}
\caption{\chandra\ images of SNR RX\,J1713.7$-$3946 \cite{Uchi07}
in an energy interval of 1--2.5 keV (panels {\bf a} and {\bf b}) or 3.5--6 keV (panel {\bf c}). In panel {\bf a}, the \hess\ contours ($> 0.7$ TeV) are overlaid on the \chandra\ map. 
A sequence of X-ray observations in July 2000, July 2005 and May 2006 for a small box depicted in panel {\bf a} are shown in panels {\bf b} and {\bf c}, 
which demonstrates time variability of synchrotron X-ray emission.}
\label{fig:RXJ1713}
\end{figure}
%%%%%%%%%%%%%%%%%%%%%%%%%%%%%%%%%%%

Theoretically it has been proposed that a turbulent magnetic field can be significantly amplified by a CR current driven instability (and other instabilities) in the shock precursor ahead of a shock front \cite{Bell04}. Magnetic field amplification (MFA) is now considered to be the key element in non-linear DSA theory \cite{Bykov11_MFA_Review}. Modeling of a synchrotron component depends critically on MFA. Consequently, any attempts to disentangle $\gamma$-ray emission mechanisms depend on our understanding of MFA, which in turn manifests in the synchrotron X-ray data. 

\subsection{Synchrotron X-ray Emission as a Probe of Magnetic Turbulence}
\label{sec:Bykov}

During the DSA process, energetic particles have to be efficiently scattered by magnetic fluctuations in the shock vicinity to be accelerated to very high energies. In addition to the strength of the magnetic field, a power spectrum of magnetic turbulences is a key parameter in the DSA theory. 

The turbulent magnetic fields, amplified possibly by CR current driven instabilities, can be imprinted in synchrotron X-ray images \cite{Bykov09}. The synchrotron emissivity depends strongly on the local magnetic field as $B^{(\Gamma +1)/2}$ (where $\Gamma$ denotes the effective photon index). Therefore, localized non-steady magnetic field concentrations contribute significantly to the synchrotron X-ray emission by the highest energy electrons in the cut-off region of the electron distribution \cite{Bykov09}. Strong fluctuations of the magnetic fields result in an intermittent, twinkling appearance of synchrotron X-ray images even if the electron distribution is steady. This may explain the variable filamentary and clumpy structures in the synchrotron X-ray map of SNR RX\,J1713.7$-$3946 (Figure\,\ref{fig:RXJ1713}). Indeed, \suzaku\  broadband X-ray observations have shown that the X-ray emission is formed in the cut-off region of the electron distribution \cite{Tanaka08}.
%Simulated synchrotron X-ray intensity and polarization maps are shown in Figure\,\ref{fig:Bykov} \cite{Bykov09}. 
%The temporal evolution of the synchrotron X-ray images is different at different photon energies. Therefore, X-ray and hard X-ray observations can verify (or discard) this scenario. 
Since the synchrotron filaments are expected to be more variable at higher photon energies in this scenario, sensitive hard X-ray imaging with \nustar\ and \astroh\ is particularly interesting. Hard X-ray observations will allow the study of the power spectra of magnetic fluctuations and the acceleration mechanisms of CRs. 
%Moreover, these compact features should be highly polarized ($\Pi \sim 50\,\%$) and therefore X-ray polarimetry (e.g., with \gems) may be able to measure the polarization. 

The important aspect of measuring $\gamma$-ray spectra with CTA will be to gauge the maximum energy 
of CR particles, particularly that of protons which, unlike electrons, do not suffer from radiative losses in SNRs. Theoretically, 
the maximum proton energy is expected to be controlled mainly by MHD waves in the shock precursor.  
As discussed above, X-ray observations provide information about such turbulent waves. 
The synchrotron X-ray stripes seen in Tycho's SNR (Figure\,\ref{fig:Eriksen}) are also intriguing in this regard; 
the pattern of the stripes may reflect a turbulent wave spectrum  \cite{Bykov11}. 
In the CTA era, a synergy between X-ray and TeV $\gamma$-ray observations of SNRs will be even greater.

\section{Gamma-ray Binaries}

\subsection{Gamma-ray Loud X-ray Binaries}

Detections of modulated TeV $\gamma$-ray emission synchronized with an orbital period  from a few high-mass X-ray binaries have revealed the existence of $\gamma$-ray loud X-ray binaries, so-called gamma-ray binaries. Examples include LS\,5039 \cite{HESS06_LS5039}, PSR\,B1259$-$63 \cite{HESS05_PSRB1259}, and LS\,I$+61^\circ 303$ \cite{albert08}, which have been well studied in the X-ray and TeV bands. Recent GeV $\gamma$-ray observations of three binaries \cite{LAT09_LS5039,LAT09_LSI,LAT11_PSRB1259} with the \lat\ demonstrate that they are also luminous GeV sources; the GeV $\gamma$-ray energy flux exceeds both the X-ray and TeV $\gamma$-ray energy fluxes at least at some orbital phase. Gamma-ray binaries can serve as an excellent laboratory for the study of extreme particle acceleration in a periodically changing environment in the vicinity of a massive star.

Two major competing models of nonthermal emission have been discussed in the literature. The first one attributes the high-energy phenomena to the interactions of a young rotation-powered pulsar with the wind (or disk) of a companion star. Collisions between the pulsar's relativistic wind and the stellar wind lead to the formation of a {\it compactified pulsar wind nebula} (CPWN), a scaled-down version of pulsar wind nebulae \cite{TA97}.  
Another model invokes {\it a microquasar}; a relativistic jet ejected by an accreting compact object accounts for the $\gamma$-ray loudness in this case \cite{Bosch08}. 

The emission from the PSR\,B1259$-$63 binary is powered by a young non-accreting pulsar and this is a clear example of the CPWN system.
Cygnus X-3 has been known as a microquasar and the GeV $\gamma$-rays detected by the \lat\ \cite{LAT09_CygX3} and \agile\ \cite{AGILE09_CygX3} can be ascribed to the emission from a relativistic jet.
However, type classification is difficult in most cases (e.g., LS\,5039). 
In what follows, we discuss two TeV-$\gamma$-ray-emitting binaries, PSR\,B1259$-$63 and LS\,5039. 

\subsection{PSR\,B1259$-$63}

PSR\,B1259$-$63 is a young radio pulsar (spin period 48\,ms) orbiting a fast-rotating O-type star LS\,2883 \cite{Negu11} in a highly eccentric 3.4\,yr orbit. The spindown power of the pulsar is $\dot{E}_{\rm p} \simeq 8\times 10^{35}$\,erg\,s$^{-1}$. Figure\,\ref{fig:LC_PSRB1259} shows the orbital lightcurves of the PSR\,B1259$-$63 system in the X-ray, GeV/TeV $\gamma$-ray, and radio bands \cite{LAT11_PSRB1259}.

%%%%%%%%%%%%%%%%%%%%%%%%%%%%%%%%%%%
\begin{figure}
\begin{center}
\includegraphics*[scale=0.425]{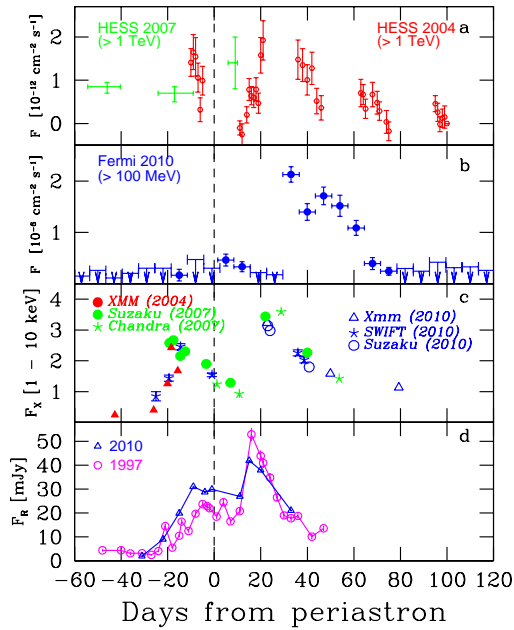}
\end{center}
\caption{Lightcurves of the PSR\,B1259$-$63 system in the (a) TeV (\hess), (b) GeV (\lat) (c) X-ray and (d) radio (2.4 GHz) bands \cite{LAT11_PSRB1259}. X-ray fluxes are in units of $\rm erg\ cm^{-2}\ s^{-1}$.}
\label{fig:LC_PSRB1259}
\end{figure}
%%%%%%%%%%%%%%%%%%%%%%%%%%%%%%%%%%%

The X-ray peaks of pre- and post-periastron are thought to arise from the interactions between the pulsar wind and the equatorial disk of the optical star. The narrow-band X-ray spectrum of the system is characterized by a power law of highly variable photon index $\Gamma \simeq 1.2 \mbox{--} 2.0$ without any detectable line emission \cite{Chern09,Uchi09}. A spectral break around $\varepsilon_{\rm br} \sim 5$\,keV was found by \suzaku\ during the pulsar's transit of the disk, which provides an important constraint on the models \cite{Uchi09}.

The electrons and positrons are presumably accelerated at an inner shock front of the pulsar wind  and adiabatically expand in the relativistic flow of the pulsar cavity. Synchrotron radiation by the accelerated electrons offers a reasonable explanation for the observed X-ray emission. The TeV $\gamma$-ray emission can be understood in terms of the anisotropic IC scattering on the intense stellar photons of the same population of electrons and positrons that produce the X-ray emission \cite{Khan07}. Future sensitive TeV observations with \cta\ will allow detailed investigations of the X-ray and TeV connection. 

Recent \lat\ observations have detected a remarkable GeV flare from PSR\,B1259$-$63 about 10 days after the second X-ray peak \cite{LAT11_PSRB1259}. The GeV luminosity reaches a sizable fraction of the pulsar's spindown power, implying a very efficient ($\sim 100\%$) conversion of the kinetic energy of the wind into $\gamma$-radiation. An extrapolation of the X-ray spectrum smoothly connects with the flare spectrum, suggesting that  the GeV $\gamma$-ray emission might be a tail of the synchrotron spectrum. This requires extremely fast acceleration of electrons and positrons, as in the case of the Crab Nebula's GeV flare. Alternatively, the GeV flare may be explained by Comptonization of the cold pulsar wind with a wind Lorentz factor of $\gamma_{\rm w} \sim  10^4$ \cite{Khan11}.

\subsection{LS\,5039}

LS\,5039 is a high-mass X-ray binary with extended radio emission, comprised of a massive O-type star and a compact object (either neutron star or black hole). A periodic TeV $\gamma$-ray signal modulated with an orbital period of 3.906 days has been detected by \hess\ \cite{HESS06_LS5039}, as shown in Figure\,\ref{fig:LC_LS5039}. 

%%%%%%%%%%%%%%%%%%%%%%%%%%%%%%%%%%%
\begin{figure}
\begin{center}
\includegraphics*[scale=0.64]{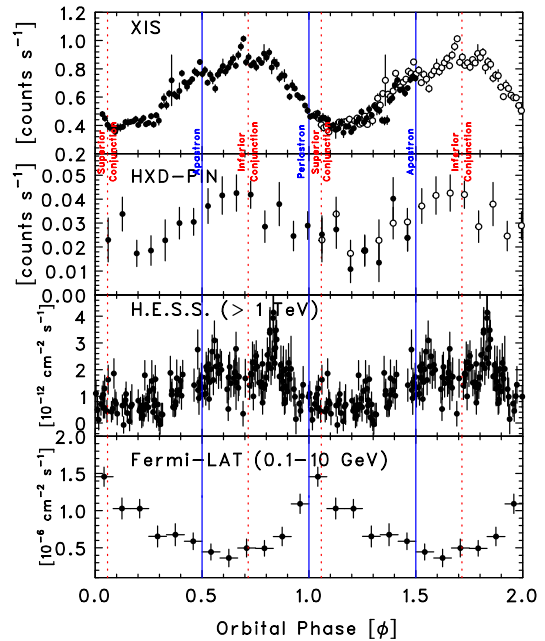}
\end{center}
\caption{Lightcurves at X-ray (\suzaku\ XIS), hard X-ray (\suzaku\ HXD), TeV (\hess), and GeV (\lat) bands of LS\,5039 as a function of orbital phase \cite{Takahashi09,LAT09_LS5039}.}
\label{fig:LC_LS5039}
\end{figure}
%%%%%%%%%%%%%%%%%%%%%%%%%%%%%%%%%%%

\suzaku\ observations, which continuously covered more than one orbital period, showed strong modulation of the X-ray emission at the orbital period \cite{Takahashi09}. In Figure\,\ref{fig:LC_LS5039}, the lightcurves obtained with the \suzaku\ XIS and HXD are compared with the $\gamma$-ray modulated curves. The X-ray spectrum measured up to 70\,keV can be described by a simple power law with a phase-dependent photon index of $\Gamma = 1.45 \mbox{--} 1.61$. A remarkably stable X-ray modulation over a long time span of $\sim 10$\,yr was revealed through a comparison with the measurements made by previous missions \cite{Kisisita}. This finding favors the CPWN nature of LS\,5039. 

The X-ray emission is likely due to synchrotron radiation, while IC scattering on stellar photons by the same population of relativistic electrons is a viable mechanism of TeV $\gamma$-ray production. The intense stellar light has dual effects; it largely enhances anisotropic IC emission but also introduces $\gamma\gamma$ opacity due to pair production, which can be used to constrain the emission site \cite{ST08}. 
While the orbital modulation of TeV $\gamma$-ray emission is affected by $\gamma\gamma$ absorption and anisotropic IC, X-rays are free of these effects so their modulation contains fundamental information about the system itself. 
For example, the X-ray lightcurve suggests the importance of adiabatic losses. 
To simultaneously explain the X-ray/TeV data,
one needs to invoke 
the extremely efficient and rapid acceleration process, allowing for acceleration of 10\,TeV electrons on a timescale of seconds.

Unlike TeV $\gamma$-rays, GeV photons are almost unaffected by $\gamma\gamma$ absorption, allowing us to probe particle acceleration in the direct vicinity of  a massive star. LS\,5039 has been detected by the \lat\ \cite{LAT09_LS5039}. Figure\,\ref{fig:LC_LS5039} shows the LAT light curve folded with the orbital period. The LAT flux peaks near periastron, which is naturally expected in the IC model. Spectral modeling suggests the presence of a second population of $e^{\pm}$ accelerated possibly in the shocked stellar wind \cite{Bednarek11}. 

Future X-ray polarimetry with GEMS may be able to confirm the synchrotron origin of the X-ray emission. 
Also, thermal X-ray emission from the shocked stellar wind, which constrains the properties of the pulsar and stellar winds, could be detectable with the \astroh\ SXS. 
Finally, thanks to \cta,  ``phase-resolved'' $\gamma$-ray spectra will be obtained in the TeV range, which would allow for an identification of the $\gamma\gamma$ absorption feature.

\section{Active Galaxies}
\label{Sec:AGN}

The phenomenon of Active Galactic Nuclei (AGN) is related to accreting supermassive black holes (SMBHs) hosted by massive galaxies. The integrated radiative output of the accreting matter in AGN dominates the extragalactic background light in the X-ray band \cite{XRB}, while the non-thermal emission of the plasma outflowing from AGN in the form of relativistic jets is widely believed to provide the bulk, or at least a substantial fraction of the extragalactic background photons at $\gamma$-ray energies \cite{GRB}. A multiwavelength approach is required for a proper understanding of AGN physics, with the X-ray and $\gamma$-ray bands being particularly important regimes to explore.

\subsection{Blazars}
\label{Sec:blazars}

Some AGN produce jets, i.e., collimated streams of magnetized plasma outflowing with relativistic bulk velocities from the immediate vicinities of SMBHs, and carrying huge amounts of energy far beyond the host galaxies \cite{JETS}. Jetted AGN observed at small viewing angles with respect to the jet axis ($< 10$\,deg) are called blazars \cite{unification}. A relatively diverse blazar family includes low-power sources of the BL Lacertae type (BL Lacs) and powerful flat-spectrum radio quasars (FSRQs). 

What is common for all the blazars is that their broad-band spectra are dominated by the non-thermal jet emission produced at parsec and sub-parsec distances from the active centers. This emission, strongly boosted in the observer rest frame due to the relativistic bulk velocities of the emitting plasma, extends from radio up to very high energy $\gamma$-ray frequencies. This is in either the X-ray or the $\gamma$-ray regime where most of the radiatively dissipated power is released. The other crucial characteristic of blazar emission is its variability, ranging from minutes to years and decades, and involving flux changes from a few percent up to a few orders of magnitude. All of these findings point toward a highly non-stationary character of AGN jets, and a very efficient and extremely rapid acceleration of the jet particles to ultrarelativistic energies, typically ascribed to Fermi-type processes at relativistic shocks.

About 100 blazars, mostly of the FSRQ type, have been associated with $\gamma$-ray sources detected by EGRET onboard \cgro\ at GeV photon energies \cite{EGRET}. Several BL Lacs have also been detected in the TeV range by the previous generation of Cherenkov telescopes, starting from Mrk\,421 \cite{Punch1992}, however with a little overlap with the EGRET catalog. After the first two years of the \lat\ operation, roughly 1,000 blazars have been identified as GeV emitters, with an almost equal split between BL Lacs and FSRQs \cite{2LAC}. In addition, the first cases of the detection of TeV flares from FSRQs were recently reported \cite{3C279,PKS1222}. Still, the overwhelming majority of the TeV blazars --- the population which nowadays is growing quite rapidly due to the development and successful operation of the modern Cherenkov telescopes --- are BL Lacs, mostly the ones selected from X-ray surveys (see, e.g., \cite{Holder}). 

%%%%%%%%%%%%%%%%%%%%%%%%%%%%%%%%%%%
\begin{figure}[t]
\begin{center}
\includegraphics[scale = 0.925]{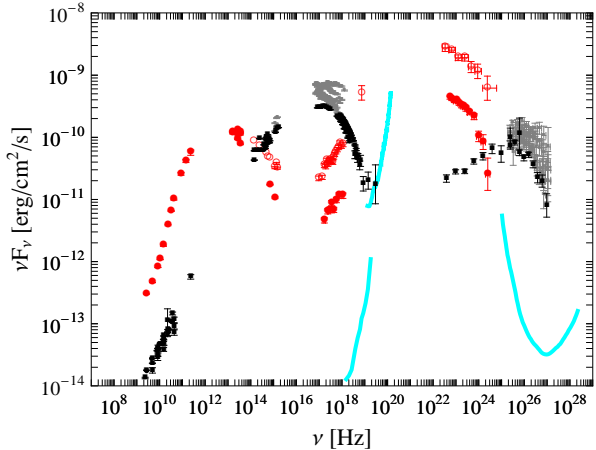}
\end{center}
\caption{Two examples of different types of blazars detected at $\gamma$-ray frequencies: 3C\,454.3 (red circles) and Mrk\,421 (black and gray squares). The broad-band spectrum of Mrk\,421 averaged over the observations taken during the 2009 multifrequency campaign and corresponding to the lower/average level of the source activity, is denoted by black squares \cite{LAT-Mrk421}. Gray squares illustrate the X-ray and TeV variability of Mrk\,421 during the 2008 higher active state \cite{MAGIC-Mrk421}. The quasi-simultaneous 2008 data for 3C\,454.3 denoted by red filled circles are taken from \cite{LAT-3C454}. The simultaneous observations of 3C\,454.3 during its flaring state (2009 Dec 3), plotted as red open circles, are analyzed and discussed by \cite{Bonnoli2011}. The thick cyan curves represent the simulated continuum sensitivities of the \astroh\ HXI and SGD instruments for point sources and 100\,ks exposures, as well as the \cta\ sensitivity for 50\,h exposure at a zenith angle of $20$\,deg with the candidate configuration C \cite{CTA}.}
\label{fig-blazars}
\end{figure}
%%%%%%%%%%%%%%%%%%%%%%%%%%%%%%%%%%%

Figure\,\ref{fig-blazars} presents the broad-band spectral energy distributions (SEDs) of particularly bright examples of the main two types of blazars, namely of Mrk\,421, which is a famous BL Lac object, and 3C\,454.3, which is an archetypal FSRQ. The main feature to notice in the figure is that the non-thermal continua consist of two highly variable spectral components; this is a generic property of blazar spectra. The low-energy component, peaking in the $\nu - \nu F_{\nu}$ representation in infrared for 3C\,454.3, and in X-rays for Mrk\,421, is established to be due to the synchrotron emission of ultrarelativistic electrons. The high-energy component, peaking in the $\gamma$-ray regime in all cases, is most successfully modeled in terms of the inverse-Compton (IC) emission of the same population of electrons \cite{Maraschi1992,Sikora1994}. 

An alternative interpretation of the high-energy blazar component, dealing with the interactions of ultrarelativistic protons with background electromagnetic fields (photomeson production, proton synchrotron emission, and the related cascades of the secondary particles) remains however a formal possibility \cite{Mannheim1993,Aharonian2000}. One of the main scientific objectives of the modern Cherenkov telescopes and \lat\ is in fact to distinguish between the two scenarios by providing conclusive evidence for the leptonic or hadronic origin of the detected $\gamma$-rays.

Interestingly, crucial pieces of information in the debate on the origin of the high-energy emission of blazars are gathered by means of X-ray observations. In the case of BL Lacs, the X-ray domain probes the highest-energy electrons ($E_e \geq 1$\,TeV) which, in the framework of the leptonic scenario, also produce the TeV photons via the synchrotron self-Compton process \cite{Tavecchio1998}. The correlated variability in the X-ray and TeV bands established for many BL Lacs and Mrk\,421 in particular \cite{Takahashi2000,Fossati2008,MAGIC-Mrk421}, therefore provides strong support for the IC origin of the observed $\gamma$-rays (see Figure\,\ref{fig-fossati}). Yet the exact correlation patterns emerging from detailed analysis revealed at the same time a picture which is much more complex than that expected in the ``standard'' models, assuming a single homogeneous emission zone and a simplified prescription for the shock acceleration of the radiating particles.

Possibly the most surprising of all such results is related to the BL Lac object PKS\,2155$-$304, for which the Cherenkov telescope \hess\ detected order-of-magnitude flares at TeV energies with doubling timescales as short as 200\,s, accompanied by only modest flux enhancement at X-ray photon energies \cite{PKS2155a}. This discovery has led to many question regarding the structure of the blazar emission zone and the particle acceleration processes involved \cite{Begelman2008}. 

%%%%%%%%%%%%%%%%%%%%%%%%%%%%%%%%%%%
\begin{figure}
\begin{center}
\includegraphics[scale = 0.3,angle = 270]{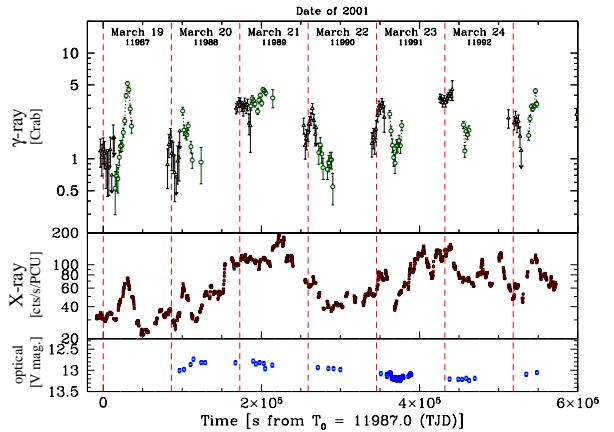}
\end{center}
\caption{Simultaneous optical ($V$ band, bottom), X-ray ($2-10$\,keV, middle) and TeV $\gamma$-ray (top) light curves for Mrk\,421 for the March 18-25, 2001 period (from \cite{Fossati2008}).}
\label{fig-fossati}
\end{figure}
%%%%%%%%%%%%%%%%%%%%%%%%%%%%%%%%%%%

To understand the multiwavelength correlations as well as the extremely short variability timescales characterizing BL Lacs, more data and higher-quality data are needed, and these can be gathered only by means of truly simultaneous, truly multiwavelength and long-term campaigns which are focused not exclusively on flaring activity of the targets, but also on their low-flux-level quiescent states. Such campaigns were hardly possible in the past due to the limited sensitivity of the previously available instruments. Recently, in the \emph{Fermi} era, the situation has improved \cite{LAT-Mrk421,LAT-Mrk501}, but still not many BL Lacs are bright enough (especially during their quiescent states) for the currently operating high-energy instruments to extract series of high-quality spectra in short multiple exposures.

The participation of X-ray telescopes providing detailed spectral information in future monitoring programs involving \cta\ is crucial because, as stated above, the X-ray domain carries information regarding the electrons directly involved in shaping the $\gamma$-ray properties of BL Lacs. Instruments such as HXI and SGD onboard \astroh\ are expected to be particularly relevant, since these will enable us, for the very first time, to track the spectral evolution of several bright BL Lacs above 10\,keV photon energies in $\sim 100$\,ks (or shorter) exposures. An exciting and unique possibility is the detection of the polarization in hard X-rays by SGD during particularly strong flaring states of the brightest BL Lacs, like that of Mrk\,501 in 1997 when the synchrotron continuum of the source extended up to the $\gtrsim 100$\,keV photon energy range \cite{Pian1998}.

A precise characterization of the X-ray spectra of TeV-emitting BL Lacs is of importance also for another reason: the highest energy $\gamma$-ray photons emitted by cosmologically distant sources suffer from the absorption by the extragalactic background light (EBL) in the IR--to--UV band due to the photon-photon annihilation process \cite{Gould1966}. The observed TeV blazar fluxes have to be therefore de-absorbed --- with the correct number density of the EBL photons for a given redshift of a target --- before attempting any modeling. But the exact spectral distribution of the EBL, which is shaped by the cosmological evolution of galaxies, is not known precisely \cite{Hauser2001}. One of the main science goals of \cta\ is, in fact, to invert the problem and to constrain the evolution of the EBL by a precise characterization of the blazar spectra enabling reliable identification and analysis of the EBL-related absorption features in the TeV range. With no good-quality and simultaneous broad-band X-ray data, on the other hand, disentangling the intrinsic curvatures and the absorption features in the observed $\gamma$-ray spectra of blazars relies heavily on several assumptions regarding the energy distribution of the emitting electrons \cite{HESS-EBL}. Any more robust determination of the intrinsic TeV properties of distant BL Lacs, and so any precise determination of the EBL level, requires simultaneous high-quality X-ray observations. 

In the case of FSRQs the situation is different than in the case of BL Lacs, since here the X-ray observations probe instead the low-energy side of the high-energy emission component (see Figure\,\ref{fig-blazars}). This high-energy component in the spectra of FSRQs dominates energetically over the synchrotron one, reaching apparent $\gamma$-ray luminosities as large as $\sim 10^{49}$\,erg\,s$^{-1}$ during the flaring states \cite{Bonnoli2011}. FSRQs display in addition very flat X-ray continua, in many cases characterized by photon indices $\Gamma_{\rm X} < 1.5$ within a broad energy range from below keV up to hundreds of keV \cite{Sikora2009}. 

In the framework of the leptonic models, the low-energy tail of the high-energy emission components of FSRQs is produced via the IC process involving the lowest-energy electrons, down to the mildly-relativistic regime \cite{Ghisellini2009}. A large amount of such mildly-relativistic leptons, outnumbering the ultrarelativistic electron population and carrying the bulk of the total jet kinetic power, should however manifest as a distinct steep-spectrum component in soft X-rays. The fact that the X-ray continua of FSRQs are flat and extend as such down to the lowest X-ray frequencies therefore has important implications for the jet energetics: as demonstrated by several authors (e.g., \cite{Sikora2000,Celotti2008}) the lack of any pronounced soft X-ray excess in the spectra of FSRQs excludes in particular the case of particle-dominated purely leptonic jets, implying either significant amount of protons, \emph{or} Poynting flux-dominated outflows. 

A caution here is that the above conclusion is based on possibly oversimplified emission models, which recently have been questioned to some extent by the aforementioned detections of short TeV flares from several FSRQs (e.g., \cite{Tavecchio2011}). More extensive $\gamma$-ray monitoring of FSRQs with continuous coverage at X-ray photon energies involving soft and hard X-ray instruments like SXS, HXI and SGD onboard the \astroh\ is therefore needed for a robust determination of the jet energetics in the systems. We also note that the hard X-ray regime is particularly well suited for studying high-redshift FSRQs, which are of interest for understanding the cosmological evolution of jetted AGN \citep{Volonteri2011}.

\subsection{Radio Galaxies}
\label{Sec:RGs}

Radio galaxies (RGs), with their relativistic jets oriented at intermediate and larger viewing angles with respect to the line of sight, are believed to constitute the parent population of blazar sources \cite{unification}. As a result of larger inclinations, the observed non-thermal emission produced within the innermost parts of the jets in RGs is not amplified by relativistic beaming as dramatically as in the case of blazars, and hence different emission components, which are hardly observable in blazar spectra, may become prominent. For RGs oriented at particularly large inclinations, the radiative output of unresolved jets may be even strongly de-beamed in the observer rest frame, and therefore bulk of the observed emission may originate at further distances from the nuclei where relativistic outflows decelerate substantially so that beaming effects become less severe.

Before the launch of \lat\ only one radio galaxy, Cen\,A, has been firmly established as a source of the MeV--GeV photons by \cgro\ \cite{CGRO-CenA}. At higher photon energies, longer exposure by the Cherenkov telescope \hegra\ allowed for the tentative detection of another object, M\,87 \cite{HEGRA-M87}. Both sources are low-power but particularly nearby systems. These two cases, when compared with about 100 blazars detected by EGRET and the previous Cherenkov telescopes, imply that RGs are relatively weak $\gamma$-ray emitters. Yet they are not ``$\gamma$-ray silent''. Indeed, after two years of the \lat\ operation and with the new generation of Cherenkov telescopes in hand, the sample of RGs detected at $\gamma$-rays has increased up to about 10 targets in the GeV range \cite{LAT-MAGN,2LAC}, and four objects at TeV photon energies \cite{VHE-CenA,VHE-IC310}. \cta\ will hopefully further enlarge the population of non-blazar TeV-emitting AGN. 

We also mention here that recently \lat\ has resolved giant (Mpc-scale) lobes surrounding the Cen\,A radio galaxy at GeV energies \cite{LAT-CenA-lobes}, proving in this way that $\gamma$-rays are being efficiently generated there, despite the advanced age and relaxed nature of the structure (see Figure\,\ref{fig-CenA-lobes}, and \cite{Takeuchi2012} for the case of the giant radio galaxy NGC\,6251). Probing the Cen\,A lobes at TeV and X-ray photon energies awaits future observations.

%%%%%%%%%%%%%%%%%%%%%%%%%%%%%%%%%%%
\begin{figure}
\begin{center}
\includegraphics[scale = 1.0]{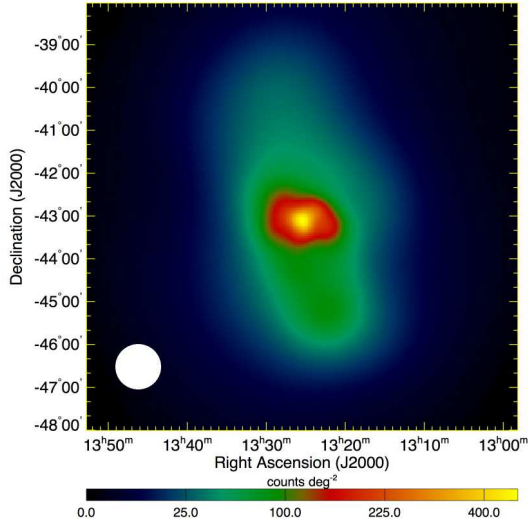}
\end{center}
\caption{Adaptively smoothed \lat\ $\gamma$-ray ($>200$\,MeV) counts maps centered on Cen\,A radio galaxy (from \cite{LAT-CenA-lobes}), showing emission from the extend lobes in the system.}
\label{fig-CenA-lobes}
\end{figure}
%%%%%%%%%%%%%%%%%%%%%%%%%%%%%%%%%%%

Increasing the sample of ``$\gamma$-ray loud'' RGs is important for several reasons. First, modeling of such sources provides an independent check of blazar models which are being developed, since, as noted above, RGs should be considered as blazars observed at larger viewing angles. Second, as also already emphasized, $\gamma$-ray observations of RGs may reveal some ``exotic'' or at least non-standard processes possibly related to the production of high energy photons and particles within active nuclei \citep{Neronov2007}, large-scale jets \cite{Stawarz03} and extended lobes \cite{Hardcastle-CenA}. Third, increasing the sample of $\gamma$-ray RGs will enable us to understand the contribution of nearby non-blazar AGN to the extragalactic $\gamma$-ray background \cite{Inoue2011}. And fourth, as discussed below in more detail, studying RGs at $\gamma$-ray frequencies may shed some light on the still hardly understood jet launching mechanisms in AGN. All of these research directions require a multiwavelength approach, but the relevance of the joint X-ray observations is particularly obvious in the latter case. 

Similarly as in the case of blazars, the gathered X-ray data for RGs offer, in principle, constraints on the highest-energy and lowest-energy segments of the population of the radiating non-thermal electrons (see \S\,\ref{Sec:blazars}). But unlike in blazar sources, the observed X-ray emission of RGs is significantly contributed, or sometimes even entirely dominated by the emission of the accretion disks and disk coronae \cite{Antonucci2011}. This constitutes an opportunity to investigate the jet-disk coupling, and hence the jet launching processes, by means of joint X-ray and radio or $\gamma$-ray observations of RGs.

%%%%%%%%%%%%%%%%%%%%%%%%%%%%%%%%%%%
\begin{figure}[t]
\begin{center}
\includegraphics[scale = 0.925]{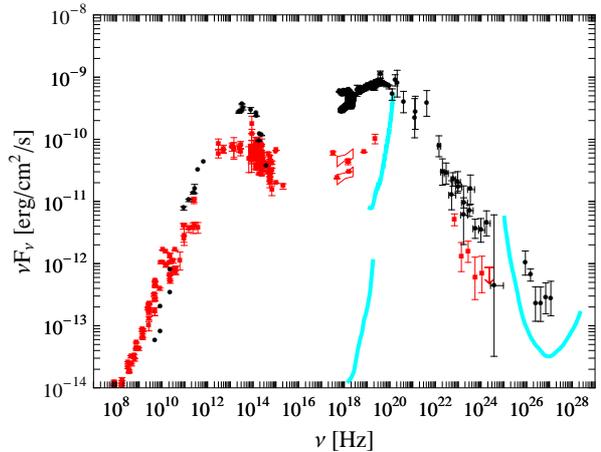}
\end{center}
\caption{Two examples of different types of radio galaxies detected at $\gamma$-ray frequencies: Cen\,A (black circles) and 3C\,120 (red squares). The compiled historical data corresponding to the unresolved core of a low-power but nearby galaxy Cen\,A, including more recent LAT and \hess\ detections, are taken from \cite{LAT-CenA-core}. The archival non-simultaneous data for the unresolved core of a high-power galaxy 3C\,120, including reanalyzed LAT fluxes (following \cite{kataoka-BLRGs}), are represented by red squares. The sensitivity curves for different X-ray and $\gamma$-ray instruments are the same as in Figure\,\ref{fig-blazars}.}
\label{fig-RGs}
\end{figure}
%%%%%%%%%%%%%%%%%%%%%%%%%%%%%%%%%%%

In Figure\,\ref{fig-RGs} we plot the broad-band SEDs of two RGs detected at $\gamma$-ray frequencies: the low-power system Cen\,A and the high-power galaxy 3C\,120. In the case of Cen\,A the multiwavelength spectrum of the unresolved core seems to be dominated by the blazar-type emission of a misaligned jet, and in particular the high-energy emission component, extending up to the TeV range, seemingly resembles the high-energy IC peak of blazar sources. The most recent but ``standard'' blazar-type modeling of this component, although quite successful up to GeV frequencies, can however hardly accommodate the observed TeV fluxes \cite{LAT-CenA-core}. Hence a contribution from some other ``exotic'' processes at the highest photon energies, or modification of the blazar modeling, are implied. To make the situation more complex, there is an ongoing debate on whether the X-ray continuum of the Cen\,A core is indeed entirely due to the jet rather than the accretion flow, and what is the contribution of the extended structures in the source to the observed TeV emission.

The observations of nearby RGs with future X-ray telescopes such as \gems, \nustar\ and \astroh\ will enable the jet and the disk contributions to the observed X-ray emission of the systems to be disentangled by obtaining high-quality spectra with unprecedented energy resolution up to tens and hundreds of keV photon energies. Such a rich and previously hardly available dataset, in addition to providing important diagnostics regarding the accretion process in AGN, will also allow us to understand the origin of $\gamma$-ray emission of RGs.

3C\,120 constitutes a particularly interesting case in this respect. Here the entire X-ray emission was argued to be produced by the accreting matter, with the possible exception of the soft X-ray band \cite{Kataoka2007}. Variable GeV emission of the source, on the other hand, tentatively detected by LAT, seems to be related to the pc-scale jet \cite{kataoka-BLRGs}. Importantly, the long-term monitoring of 3C\,120 at X-ray and radio frequencies has revealed a nontrivial connection between the two bands, with the dips in the X-ray emission followed by ejections of bright superluminal knots along the radio outflow \cite{Marscher2002}. In this way a direct observational link between the accretion and jet launching processes has been established for the very first time in the case of an AGN. The proposed interpretation involved X-ray dips due to the disappearance of the inner parts of the accretion disk leading to the ejection of the excess matter along the jet axis, analogous to what is observed in the Galactic jet sources. The possibility that the related phenomenon may be also observed at $\gamma$-ray frequencies, with the $\gamma$-ray flares following the dips in the accretion-related X-ray emission, awaits the operation of \cta\ and future X-ray missions.

\section{Clusters of Galaxies}
\label{Sec:Clusters}

Merging processes leading to the formation of clusters of galaxies release huge amounts of gravitational energy ($\gtrsim 10^{64}$\,erg) on timescales of the order of $\sim$\,Gyr \cite{Sarazin1986}. While much of this energy is contained in thermal plasma with temperatures $kT \lesssim 10$\,keV emitting X-ray photons via the bremsstrahlung process, part of it may be channelled to accelerate a small fraction of particles from the thermal pool to ultrarelativistic energies, and to form in this way an energetically relevant population of cosmic rays (CRs) within the intracluster medium (ICM). In addition to the thermal and non-thermal baryonic particles, a large amount of dark matter (DM) is believed to be present in massive clusters of galaxies. 

Both DM and CRs are supposed to give observable signatures at $\gamma$-ray frequencies, due to DM annihilation or decay processes, and due to interactions of hadronic CRs with the ambient gas \cite{Blasi2007,Jeltema2009}. There is an ongoing search for such signatures using currently available $\gamma$-ray instruments. Importantly, however, the production of $\gamma$-rays by DM and CRs should be accompanied by the production of secondary $e^{\pm}$ pairs. The presence of such secondary leptons should then result in observational signatures at lower frequencies, and in particular in the radio and X-ray domains.

Non-thermal activity in the ICM manifests clearly in the phenomenon of giant ($\sim$\,Mpc-scale) radio halos. These are roughly spherical and low-surface brightness structures centered at the position of the peaks in galaxy distributions, which are found in about $10\%$ of the systems \cite{Ferrari2008}. It was long speculated whether the synchrotron-emitting electrons populating giant halos are in fact secondary particles resulting from hadronic CR interactions \cite{Dennison1980}, or even DM annihilation/decay processes. Even though this possibility still cannot be excluded, a number of arguments and considerations were presented against such a scenario \cite{Magic-clusters}. Instead, radio-emitting electrons forming giant radio halos are now most widely believed to be accelerated directly from the thermal pool of the ICM by magnetic turbulence induced by merger processes at relatively early stages of the cluster lifetime \cite{Petrosian2001}.

%%%%%%%%%%%%%%%%%%%%%%%%%%%%%%%%%%%
\begin{figure}[t]
\includegraphics[scale = 0.925]{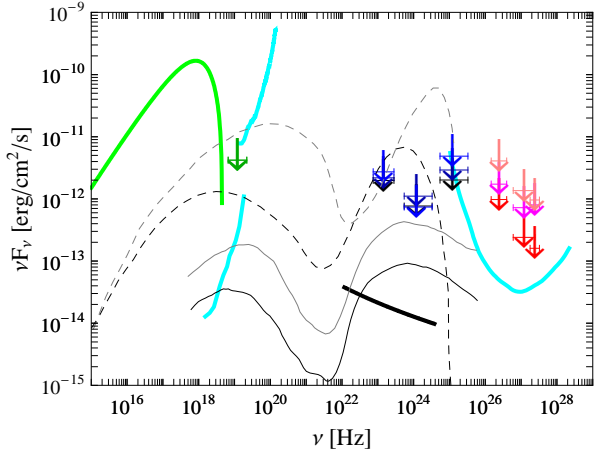}
\caption{High to very high-energy pectrum of the Coma cluster of galaxies (Abell 1656). Thick green curve illustrates the thermal emission of the cluster gas ($k T \simeq 8.3$\,keV, $L_{\rm 0.1-2.4\,keV} \simeq 3 \times 10^{44}$\,erg\,s$^{-1}$). Hard X-ray ($20-80$\,keV) upper limit denoted by dark green arrow is taken from \cite{Wik2011}. \lat\ upper limits within the $0.2-1$, $1-10$, and $10-100$\,GeV photon energy ranges corresponding to a point source, to a King profile, and to the two-dimensional Gaussian with the $68\%$ contamination radius of $0.8$\,deg centered at the position of the cluster are denoted by black, dark blue, and blue arrows, respectively \cite{LAT-clusters}. Upper limits at $1$, $5$, and $10$\,TeV photon energies for the source region of interest with the radius of 0, $0.2$ and $0.4$\,deg, as reported by the \hess\ Collaboration \cite{HESS-Coma}, are denoted by red, magenta, and pink arrows, respectively. The sensitivities for \astroh\ and \cta\ (thick cyan curves) are the same as in Figure\,\ref{fig-blazars}. The black and gray solid curves represent the predictions regarding the non-thermal emission of primary and secondary electrons accelerated by the magnetic turbulence within the Coma cluster for the central magnetic field intensity $5$\,$\mu$G and $2$\,$\mu$G, respectively (from \citep{Brunetti2011}). The black and gray dashed curves illustrate two exemplary models for the DM-induced emission in the Coma cluster for the intermediate neutralino mass of $60$\,GeV (from \citep{Colafrancesco2011}). The thick black line corresponds to the hadronic model of radio halo in the Coma cluster by \citep{Pinzke2010} normalized to the minimum flux prediction of \cite{Pfrommer2008} for the spectral index $2.1$.
}
\label{fig-Coma}
\end{figure}
%%%%%%%%%%%%%%%%%%%%%%%%%%%%%%%%%%%

It was recognized early on that the same electrons which produce synchrotron photons at radio frequencies should also lead to the production of higher-energy emission via IC upscattering of the cosmic microwave background radiation \cite{Rephaeli1979}. This additional emission component could then be probed at hard X-ray photon energies as an excess over the thermal (free-free) emission of the hot intracluster gas. Looking for such an excess using different X-ray instruments has resulted in contradicting claims in the past \cite{Rephaeli2008}. The most recent studies using \suzaku\ and \swift\ satellites indicate however that there is no power-law excess within the $10-100$\,keV photon energy range down to the level of a few $\times 10^{-12}$\,erg\,cm$^{-2}$\,s$^{-1}$ for a number of the brightest systems, with the exception of the peculiar Bullet cluster \cite{Ajello2009,Wik2011}. These upper limits, when combined with the radio data, translate into lower limits for the magnetic field intensity within the ICM $> 0.3$\,$\mu$G \cite{Rephaeli2008,Wik2011}.

Searching for the observable signatures of the hadronic CR and DM populations in galaxy clusters is therefore a multiwavelength effort, even though it predominantly involves $\gamma$-ray instruments. As such, it will continue vigorously in the future with CTA. At present, despite extensive investigation, only upper limits for the $\gamma$-ray emission of clusters of galaxies have been provided both in the GeV \cite{Reimer2003,LAT-clusters} and in the TeV ranges (e.g., \cite{HESS-Coma,Magic-clusters}), and these are typically at the level of $\sim 10^{-12}$\,erg\,cm$^{-2}$\,s$^{-1}$ (see Figure\,\ref{fig-Coma} for the case of the Coma cluster). Such limits provide constraints on the DM models, and also limit the contribution of the hadronic CRs to the cluster pressure down to the level of a few percent, at most. 

%%%%%%%%%%%%%%%%%%%%%%%%%%%%%%%%%%%
\begin{figure*}
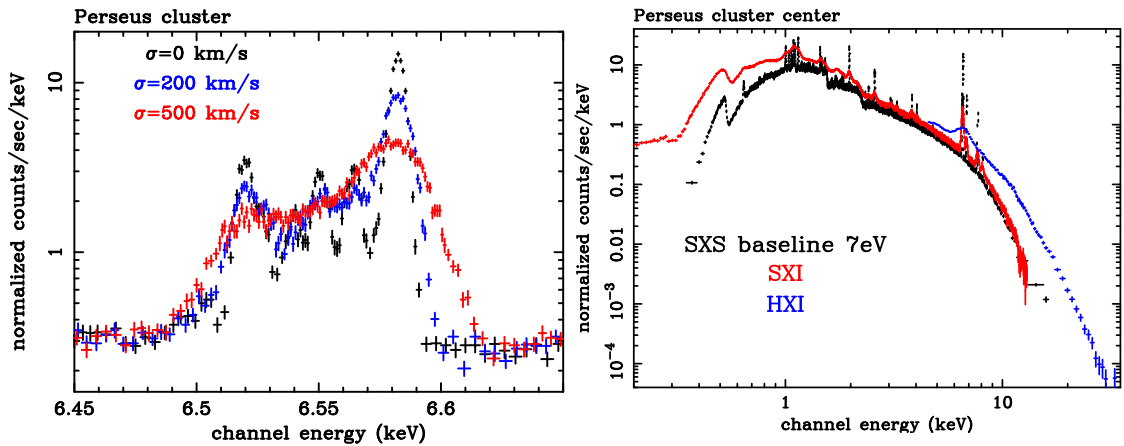

\begin{center}
\includegraphics[scale = 0.335, angle = 270]{A426_center_FeHeK.ps}
\includegraphics[scale = 0.3, angle = 270]{a426_center_sxs_sxi_hxi.ps}
\end{center}
\caption{Simulated spectra for 100\,ks \astroh\ observations of Perseus Cluster. {\bf (left)} 
SXS spectra around the iron K line complex.  
Line profiles assuming $\sigma=0$, 100 and 200\,$\rm km\ s^{-1}$ turbulence. 
{\bf (right)} SXS (black), SXI (red), and HXI (blue) spectra 
for hot plasma with three different temperatures of 0.6, 2.6 and 6.1\,keV ($r < 2'$; from \cite{Ref:ASTRO-H}).}
\label{fig-perseus}
\end{figure*}
%%%%%%%%%%%%%%%%%%%%%%%%%%%%%%%%%%%

A variety of choices regarding the candidates for the DM particles results however in different expectations regarding the high-energy emission of clusters related to the DM decay and annihilation processes \cite{Colafrancesco2011,Pinzke2011}. A variety of viable CR acceleration processes also plays a role. For example, the most widely considered acceleration mechanism for hadronic CRs in galaxy clusters is related to the 1st-order Fermi process operating at the fronts of large-scale shocks formed during the cluster mergers \cite{Bykov2000,Miniati2001}. But shock-produced CRs (both primary and secondary particles) may be also efficiently re-accelerated within the ICM via stochastic interaction with magnetic turbulence \citep{Brunetti2011}. Injection of ultrarelativistic particles into the ICM may be also related to the activity of AGN located in cluster centers \citep{Timokhin2004,Fujita2007}. The insufficient knowledge regarding the kinematics of the cluster gas and the structure of the cluster magnetic fields, affects the model predictions regarding the spatial distribution and the energy spectra of the accelerated CRs, and hence the non-thermal emission of the ICM (see Figure\,\ref{fig-Coma}). 

Future X-ray observations which will allow not only for more robust constraints of the hard X-ray emission of clusters, but which will also provide a detailed insight into the kinematics of the ICM, and hence into the various energy dissipation processes involved, are therefore of primary interest. Missions such as \astroh\ can accomplish the task by means of spectrometric observations probing bulk plasma velocities and/or turbulence at a resolution corresponding to a speed of a few $\times 100$\,km/s, \emph{together} with an arcmin imaging system in the hard X-ray band with a sensitivity of orders of magnitude better than previous missions (see Figure\,\ref{fig-perseus}).

\section{Conclusions}

During the next decade, X-ray catalogues including about 200,000 clusters located out to high redshifts, together with about 3 million AGN, will be available thanks to the operation of \erosita. Around the same time, \nustar\ and \astroh, carrying high-resolution hard X-ray mirrors with unprecedented performance, will open a new chapter in the studies of high-energy radiation of astrophysical sources in the hardly explored regime from $10$\,keV up to several hundreds of keV photon energies. A soft $\gamma$-ray survey at photon energies below 1\,MeV, with a sensitivity improved by 1--2 orders of magnitude with respect to previous surveys, will become possible thanks to the SGD instruments onboard \astroh. The collected data will enable us to monitor highly variable synchrotron emission of the highest-energy electrons in blazars and Galactic binaries, to track the evolution of supermassive black holes which are heavily obscured, and in general to probe with unprecedented accuracy the accretion process in different types of AGN. The new X-ray instruments will also uniquely allow for mapping of the spatial extent of the hard X-ray emission in diffuse non-thermal structures, thus tracing the sites of particle acceleration in clusters of galaxies and SNRs. In parallel, imaging spectroscopy with the energy resolution $< 5-7$\,eV brought by the micro-calorimeter onboard \astroh\ will reveal line broadening and Doppler shifts due to turbulent or bulk velocities in such extended systems. \gems\ will perform the first sensitive X-ray polarization survey of several classes of X-ray emitting sources characterized by strong gravitational or magnetic fields.

All these breakthroughs in studying high energy phenomena in the Universe are expected to happen at the time of the operation of the Cherenkov Telescope Array. The synergy between the TeV observations with \cta\ and the X-ray observations with the future missions discussed here can hardly be overemphasized: parallel investigations in both regimes are indeed highly complementary and indispensable to understand the complex physics of the Galactic cosmic ray accelerators, active galaxies, and clusters of galaxies. This synergy regards primarily constraining particle acceleration processes, accretion in the black hole systems, and kinematics of the background plasma in various astrophysical sources of high energy emission. The cosmological context should not be forgotten either, since future X-ray missions together with \cta\ are expected to enable significant progress in understanding the origin of the high-energy cosmic background radiation, as well as the nature of dark matter particles through the study of clusters of galaxies.

\bibliographystyle{elsarticle-num}
\bibliography{ms}

\begin{thebibliography}{100}
\expandafter\ifx\csname url\endcsname\relax
  \def\url#1{\texttt{#1}}\fi
\expandafter\ifx\csname urlprefix\endcsname\relax\def\urlprefix{URL }\fi
\expandafter\ifx\csname href\endcsname\relax
  \def\href#1#2{#2} \def\path#1{#1}\fi

\bibitem{Ref:Mrk421_TT}
T.~{Takahashi}, et~al., {ASCA Observation of an X-Ray/TeV Flare from the BL
  Lacertae Object Markarian 421}, \apjl 470 (1996) L89+.

\bibitem{Ref:RXJ1713_HESS}
F.~{Aharonian}, et~al., {A detailed spectral and morphological study of the
  gamma-ray supernova remnant RX J1713.7-3946 with HESS}, \aap 449 (2006)
  223--242.

\bibitem{Ref:ASTROSAT}
P.~C. {Agrawal}, {A broad spectral band Indian Astronomy satellite Astrosat},
  Advances in Space Research 38 (2006) 2989--2994.

\bibitem{Ref:NuSTAR}
F.~A. {Harrison}, et~al., {The Nuclear Spectroscopic Telescope Array (NuSTAR)},
  in: Society of Photo-Optical Instrumentation Engineers (SPIE) Conference
  Series, Vol. 7732 of Society of Photo-Optical Instrumentation Engineers
  (SPIE) Conference Series, 2010.
\newblock \href {http://arxiv.org/abs/1008.1362} {\path{arXiv:1008.1362}}.

\bibitem{Ref:eROSITA}
P.~{Predehl}, et~al., {eROSITA on SRG}, in: Society of Photo-Optical
  Instrumentation Engineers (SPIE) Conference Series, Vol. 7732 of Society of
  Photo-Optical Instrumentation Engineers (SPIE) Conference Series, 2010.
\newblock \href {http://arxiv.org/abs/1001.2502} {\path{arXiv:1001.2502}}.

\bibitem{Ref:ASTRO-H}
T.~{Takahashi}, et~al., {The ASTRO-H Mission}, in: Society of Photo-Optical
  Instrumentation Engineers (SPIE) Conference Series, Vol. 7732 of Society of
  Photo-Optical Instrumentation Engineers (SPIE) Conference Series, 2010.
\newblock \href {http://arxiv.org/abs/1010.4972} {\path{arXiv:1010.4972}}.

\bibitem{Ref:GEMS}
K.~{Jahoda}, {The Gravity and Extreme Magnetism Small Explorer}, in: Society of
  Photo-Optical Instrumentation Engineers (SPIE) Conference Series, Vol. 7732
  of Society of Photo-Optical Instrumentation Engineers (SPIE) Conference
  Series, 2010.

\bibitem{Ref:Athena}
M.~{Bavdaz}, et~al., {ESA-led ATHENA/IXO optics development status}, in:
  Society of Photo-Optical Instrumentation Engineers (SPIE) Conference Series,
  Vol. 8147 of Society of Photo-Optical Instrumentation Engineers (SPIE)
  Conference Series, 2011.

\bibitem{Ref:LOFT}
M.~{Feroci}, et~al., {LOFT: a large observatory for x-ray timing}, in: Society
  of Photo-Optical Instrumentation Engineers (SPIE) Conference Series, Vol.
  7732 of Society of Photo-Optical Instrumentation Engineers (SPIE) Conference
  Series, 2010.
\newblock \href {http://arxiv.org/abs/1008.1009} {\path{arXiv:1008.1009}}.

\bibitem{Atwood2009}
W.~B. {Atwood}, et~al., {The Large Area Telescope on the Fermi Gamma-Ray Space
  Telescope Mission}, \apj 697 (2009) 1071--1102.

\bibitem{Carmona2011}
E.~{Carmona}, et~al., {for the Magic Collaboration}, {Performance of the MAGIC
  Stereo System}, ArXiv e-prints:\href {http://arxiv.org/abs/1110.0947}
  {\path{arXiv:1110.0947}}.

\bibitem{CTA}
T.~{CTA Consortium}, {Design Concepts for the Cherenkov Telescope Array}, ArXiv
  e-prints:\href {http://arxiv.org/abs/1008.3703} {\path{arXiv:1008.3703}}.

\bibitem{Ref:INTEGRAL_GC}
G.~{B{\'e}langer}, et~al., {A Persistent High-Energy Flux from the Heart of the
  Milky Way: INTEGRAL's View of the Galactic Center}, \apj 636 (2006) 275--289.

\bibitem{Ref:SXS}
K.~{Mitsuda}, et~al., {The high-resolution x-ray microcalorimeter spectrometer
  system for the SXS on ASTRO-H}, in: Society of Photo-Optical Instrumentation
  Engineers (SPIE) Conference Series, Vol. 7732 of Society of Photo-Optical
  Instrumentation Engineers (SPIE) Conference Series, 2010.

\bibitem{Koyama95}
K.~{Koyama}, et~al., {Evidence for shock acceleration of high-energy electrons
  in the supernova remnant SN1006}, \nat 378 (1995) 255--258.

\bibitem{Eriksen11}
K.~A. {Eriksen}, et~al., {Evidence for Particle Acceleration to the Knee of the
  Cosmic Ray Spectrum in Tycho's Supernova Remnant}, \apjl 728 (2011) L28+.

\bibitem{Ref:Balmer-dominated}
P.~{Ghavamian}, J.~M. {Laming}, C.~E. {Rakowski}, {A Physical Relationship
  between Electron-Proton Temperature Equilibration and Mach Number in Fast
  Collisionless Shocks}, \apjl 654 (2007) L69--L72.

\bibitem{Ref:Hayato}
A.~{Hayato}, et~al., {Expansion Velocity of Ejecta in Tycho's Supernova Remnant
  Measured by Doppler Broadened X-ray Line Emission}, \apj 725 (2010) 894--903.

\bibitem{Kawasaki02}
M.~T. {Kawasaki}, et~al., {ASCA Observations of the Supernova Remnant IC 443:
  Thermal Structure and Detection of Overionized Plasma}, \apj 572 (2002)
  897--905.

\bibitem{Yamaguchi09}
H.~{Yamaguchi}, et~al., {Discovery of Strong Radiative Recombination Continua
  from the Supernova Remnant IC 443 with Suzaku}, \apjl 705 (2009) L6--L9.

\bibitem{Uchi11}
Y.~{Uchiyama}, {the Fermi-LAT Collaboration}, {GeV Gamma-Rays from Supernova
  Remnants Interacting with Molecular Clouds}, in: 25th Texas Symposium on
  Relativistic Astrophysics, 2010.
\newblock \href {http://arxiv.org/abs/1104.1197} {\path{arXiv:1104.1197}}.

\bibitem{BaadeZwicky34}
W.~{Baade}, F.~{Zwicky}, {Cosmic Rays from Super-novae}, Proceedings of the
  National Academy of Science 20 (1934) 259--263.

\bibitem{MD01}
M.~A. {Malkov}, L.~{O'C Drury}, {Nonlinear theory of diffusive acceleration of
  particles by shock waves}, Reports on Progress in Physics 64 (2001) 429--481.

\bibitem{Hillas_Review05}
A.~M. {Hillas}, {TOPICAL REVIEW: Can diffusive shock acceleration in supernova
  remnants account for high-energy galactic cosmic rays?}, Journal of Physics G
  Nuclear Physics 31 (2005) 95--+.

\bibitem{DAV94}
L.~O. {Drury}, F.~A. {Aharonian}, H.~J. {Voelk}, {The gamma-ray visibility of
  supernova remnants. A test of cosmic ray origin}, \aap 287 (1994) 959--971.

\bibitem{HESS_1713}
F.~A. {Aharonian}, et~al., {High-energy particle acceleration in the shell of a
  supernova remnant}, \nat 432 (2004) 75--77.

\bibitem{HESS_1713_2}
F.~{Aharonian}, et~al., {Primary particle acceleration above 100 TeV in the
  shell-type supernova remnant RX J1713.7-3946 with deep HESS observations},
  \aap 464 (2007) 235--243.

\bibitem{2009_TeVREview}
J.~A. {Hinton}, W.~{Hofmann}, {Teraelectronvolt Astronomy}, \araa 47 (2009)
  523--565.

\bibitem{Bamba05}
A.~{Bamba}, et~al., {A Spatial and Spectral Study of Nonthermal Filaments in
  Historical Supernova Remnants: Observational Results with Chandra}, \apj 621
  (2005) 793--802.

\bibitem{VL03}
J.~{Vink}, J.~M. {Laming}, {On the Magnetic Fields and Particle Acceleration in
  Cassiopeia A}, \apj 584 (2003) 758--769.

\bibitem{Voelk05}
H.~J. {V{\"o}lk}, E.~G. {Berezhko}, L.~T. {Ksenofontov}, {Magnetic field
  amplification in Tycho and other shell-type supernova remnants}, \aap 433
  (2005) 229--240.

\bibitem{Pohl05}
M.~{Pohl}, H.~{Yan}, A.~{Lazarian}, {Magnetically Limited X-Ray Filaments in
  Young Supernova Remnants}, \apjl 626 (2005) L101--L104.

\bibitem{Uchi07}
Y.~{Uchiyama}, et~al., {Extremely fast acceleration of cosmic rays in a
  supernova remnant}, \nat 449 (2007) 576--578.

\bibitem{Bykov09}
A.~M. {Bykov}, et~al., {A model of polarized X-ray emission from twinkling
  synchrotron supernova shells}, \mnras 399 (2009) 1119--1125.

\bibitem{Bell04}
A.~R. {Bell}, {Turbulent amplification of magnetic field and diffusive shock
  acceleration of cosmic rays}, \mnras 353 (2004) 550--558.

\bibitem{Bykov11_MFA_Review}
A.~M. {Bykov}, D.~C. {Ellison}, M.~{Renaud}, {Magnetic Fields in Cosmic
  Particle Acceleration Sources}, \ssr (2011) 191--+.

\bibitem{Tanaka08}
T.~{Tanaka}, et~al., {Study of Nonthermal Emission from SNR RX J1713.7-3946
  with Suzaku}, \apj 685 (2008) 988--1004.

\bibitem{Bykov11}
A.~M. {Bykov}, et~al., {X-ray Stripes in Tycho's Supernova Remnant: Synchrotron
  Footprints of a Nonlinear Cosmic-ray-driven Instability}, \apjl 735 (2011)
  L40+.

\bibitem{HESS06_LS5039}
F.~{Aharonian}, et~al., {3.9 day orbital modulation in the TeV {$\gamma$}-ray
  flux and spectrum from the X-ray binary LS 5039}, \aap 460 (2006) 743--749.

\bibitem{HESS05_PSRB1259}
F.~{Aharonian}, et~al., {Discovery of the binary pulsar PSR B1259-63 in
  very-high-energy gamma rays around periastron with HESS}, \aap 442 (2005)
  1--10.

\bibitem{albert08}
J.~{Albert}, et~al., {Multiwavelength (Radio, X-Ray, and {$\gamma$}-Ray)
  Observations of the {$\gamma$}-Ray Binary LS I +61 303}, \apj 684 (2008)
  1351--1358.

\bibitem{LAT09_LS5039}
A.~A. {Abdo}, et~al., {Fermi/LAT observations of LS 5039}, \apjl 706 (2009)
  L56--L61.

\bibitem{LAT09_LSI}
A.~A. {Abdo}, et~al., {Fermi LAT Observations of LS I +61\,303: First Detection
  of an Orbital Modulation in GeV Gamma Rays}, \apjl 701 (2009) L123--L128.

\bibitem{LAT11_PSRB1259}
A.~A. {Abdo}, et~al., {Discovery of High-energy Gamma-ray Emission from the
  Binary System PSR B1259-63/LS 2883 around Periastron with Fermi}, \apjl 736
  (2011) L11+.

\bibitem{TA97}
M.~{Tavani}, J.~{Arons}, {Theory of High-Energy Emission from the Pulsar/Be
  Star System PSR 1259-63. I. Radiation Mechanisms and Interaction Geometry},
  \apj 477 (1997) 439--+.

\bibitem{Bosch08}
V.~{Bosch-Ramon}, {The physics of non-thermal radiation in microquasars}, ArXiv
  e-prints:\href {http://arxiv.org/abs/0805.1707} {\path{arXiv:0805.1707}}.

\bibitem{LAT09_CygX3}
{Fermi LAT Collaboration}, A.~A. {Abdo}, et~al., {Modulated High-Energy
  Gamma-Ray Emission from the Microquasar Cygnus X-3}, Science 326 (2009)
  1512--.

\bibitem{AGILE09_CygX3}
M.~{Tavani}, et~al., {Extreme particle acceleration in the microquasar
  CygnusX-3}, \nat 462 (2009) 620--623.

\bibitem{Negu11}
I.~{Negueruela}, et~al., {Astrophysical Parameters of LS 2883 and Implications
  for the PSR B1259-63 Gamma-ray Binary}, \apjl 732 (2011) L11+.

\bibitem{Chern09}
M.~{Chernyakova}, et~al., {X-ray observations of PSR B1259-63 near the 2007
  periastron passage}, \mnras 397 (2009) 2123--2132.

\bibitem{Uchi09}
Y.~{Uchiyama}, et~al., {Suzaku Observations of PSR B1259-63: A New
  Manifestation of Relativistic Pulsar Wind}, \apj 698 (2009) 911--921.

\bibitem{Khan07}
D.~{Khangulyan}, et~al., {TeV light curve of PSR B1259-63/SS2883}, \mnras 380
  (2007) 320--330.

\bibitem{Khan11}
D.~{Khangulyan}, et~al., {Post-Periastron Gamma Ray Flare from PSR B1259-63/LS
  2883 as a Result of Comptonization of the Cold Pulsar Wind}, ArXiv
  e-prints:\href {http://arxiv.org/abs/1107.4833} {\path{arXiv:1107.4833}}.

\bibitem{Takahashi09}
T.~{Takahashi}, et~al., {Study of the Spectral and Temporal Characteristics of
  X-Ray Emission of the Gamma-Ray Binary LS 5039 with Suzaku}, \apj 697 (2009)
  592--600.

\bibitem{Kisisita}
T.~{Kishishita}, et~al., {Long-Term Stability of Nonthermal X-Ray Modulation in
  the Gamma-Ray Binary LS 5039}, \apjl 697 (2009) L1--L5.

\bibitem{ST08}
A.~{Sierpowska-Bartosik}, D.~F. {Torres}, {Pulsar wind zone processes in LS
  5039}, Astroparticle Physics 30 (2008) 239--263.

\bibitem{Bednarek11}
W.~{Bednarek}, {A model for the two component {$\gamma$}-ray spectra observed
  from the {$\gamma$}-ray binaries}, \mnras 418 (2011) L49--L53.

\bibitem{XRB}
R.~{Gilli}, A.~{Comastri}, G.~{Hasinger}, {The synthesis of the cosmic X-ray
  background in the Chandra and XMM-Newton era}, \aap 463 (2007) 79--96.

\bibitem{GRB}
A.~A. {Abdo}, et~al., {Fermi-LAT Collaboration}, {The Fermi-LAT High-Latitude
  Survey: Source Count Distributions and the Origin of the Extragalactic
  Diffuse Background}, \apj 720 (2010) 435--453.

\bibitem{JETS}
M.~C. {Begelman}, R.~D. {Blandford}, M.~J. {Rees}, {Theory of extragalactic
  radio sources}, Reviews of Modern Physics 56 (1984) 255--351.

\bibitem{unification}
C.~M. {Urry}, P.~{Padovani}, {Unified Schemes for Radio-Loud Active Galactic
  Nuclei}, \pasp 107 (1995) 803--+.

\bibitem{EGRET}
R.~C. {Hartman}, et~al., {The Third EGRET Catalog of High-Energy Gamma-Ray
  Sources}, \apjs 123 (1999) 79--202.

\bibitem{Punch1992}
M.~{Punch}, et~al., {Detection of TeV photons from the active galaxy Markarian
  421}, \nat 358 (1992) 477--+.

\bibitem{2LAC}
{The Fermi-LAT collaboration}, {The Second Catalog of Active Galactic Nuclei
  Detected by the Fermi Large Area Telescope}, ArXiv e-prints:\href
  {http://arxiv.org/abs/1108.1420} {\path{arXiv:1108.1420}}.

\bibitem{3C279}
{MAGIC Collaboration}, J.~{Albert}, et~al., {Very-High-Energy gamma rays from a
  Distant Quasar: How Transparent Is the Universe?}, Science 320 (2008) 1752--.

\bibitem{PKS1222}
J.~{Aleksi{\'c}}, et~al., {MAGIC Discovery of Very High Energy Emission from
  the FSRQ PKS 1222+21}, \apjl 730 (2011) L8+.

\bibitem{Holder}
J.~{Holder}, {TeV Gamma-ray Astronomy: A Summary}, ArXiv e-prints:\href
  {http://arxiv.org/abs/1204.1267} {\path{arXiv:1204.1267}}.

\bibitem{LAT-Mrk421}
A.~A. {Abdo}, et~al., {Fermi Large Area Telescope Observations of Markarian
  421: The Missing Piece of its Spectral Energy Distribution}, \apj 736 (2011)
  131--+.

\bibitem{MAGIC-Mrk421}
J.~{Aleksic}, et~al., {Mrk 421 active state in 2008: the MAGIC view,
  simultaneous multi-wavelength observations and SSC model constrained}, ArXiv
  e-prints:\href {http://arxiv.org/abs/1106.1589} {\path{arXiv:1106.1589}}.

\bibitem{LAT-3C454}
A.~A. {Abdo}, et~al., {Early Fermi Gamma-ray Space Telescope Observations of
  the Quasar 3C 454.3}, \apj 699 (2009) 817--823.

\bibitem{Bonnoli2011}
G.~{Bonnoli}, et~al., {The {$\gamma$}-ray brightest days of the blazar 3C
  454.3}, \mnras 410 (2011) 368--380.

\bibitem{Maraschi1992}
L.~{Maraschi}, G.~{Ghisellini}, A.~{Celotti}, {A jet model for the gamma-ray
  emitting blazar 3C 279}, \apjl 397 (1992) L5--L9.

\bibitem{Sikora1994}
M.~{Sikora}, M.~C. {Begelman}, M.~J. {Rees}, {Comptonization of diffuse ambient
  radiation by a relativistic jet: The source of gamma rays from blazars?},
  \apj 421 (1994) 153--162.

\bibitem{Mannheim1993}
K.~{Mannheim}, {The proton blazar}, \aap 269 (1993) 67--76.

\bibitem{Aharonian2000}
F.~A. {Aharonian}, {TeV gamma rays from BL Lac objects due to synchrotron
  radiation of extremely high energy protons}, \na 5 (2000) 377--395.

\bibitem{Tavecchio1998}
F.~{Tavecchio}, L.~{Maraschi}, G.~{Ghisellini}, {Constraints on the Physical
  Parameters of TeV Blazars}, \apj 509 (1998) 608--619.

\bibitem{Takahashi2000}
T.~{Takahashi}, et~al., {Complex Spectral Variability from Intensive
  Multiwavelength Monitoring of Markarian 421 in 1998}, \apjl 542 (2000)
  L105--L109.

\bibitem{Fossati2008}
G.~{Fossati}, et~al., {Multiwavelength Observations of Markarian 421 in 2001
  March: An Unprecedented View on the X-Ray/TeV Correlated Variability}, \apj
  677 (2008) 906--925.

\bibitem{PKS2155a}
F.~{Aharonian}, et~al., {An Exceptional Very High Energy Gamma-Ray Flare of PKS
  2155-304}, \apjl 664 (2007) L71--L74.

\bibitem{Begelman2008}
M.~C. {Begelman}, A.~C. {Fabian}, M.~J. {Rees}, {Implications of very rapid TeV
  variability in blazars}, \mnras 384 (2008) L19--L23.

\bibitem{LAT-Mrk501}
A.~A. {Abdo}, et~al., {Insights into the High-energy {$\gamma$}-ray Emission of
  Markarian 501 from Extensive Multifrequency Observations in the Fermi Era},
  \apj 727 (2011) 129--+.

\bibitem{Pian1998}
E.~{Pian}, et~al., {BeppoSAX Observations of Unprecedented Synchrotron Activity
  in the BL Lacertae Object Markarian 501}, \apjl 492 (1998) L17+.

\bibitem{Gould1966}
R.~J. {Gould}, G.~{Schr{\'e}der}, {Opacity of the Universe to High-Energy
  Photons}, Physical Review Letters 16 (1966) 252--254.

\bibitem{Hauser2001}
M.~G. {Hauser}, E.~{Dwek}, {The Cosmic Infrared Background: Measurements and
  Implications}, \araa 39 (2001) 249--307.

\bibitem{HESS-EBL}
F.~{Aharonian}, et~al., {A low level of extragalactic background light as
  revealed by {$\gamma$}-rays from blazars}, \nat 440 (2006) 1018--1021.

\bibitem{Sikora2009}
M.~{Sikora}, et~al., {Constraining Emission Models of Luminous Blazar Sources},
  \apj 704 (2009) 38--50.

\bibitem{Ghisellini2009}
G.~{Ghisellini}, F.~{Tavecchio}, {Canonical high-power blazars}, \mnras 397
  (2009) 985--1002.

\bibitem{Sikora2000}
M.~{Sikora}, G.~{Madejski}, {On Pair Content and Variability of Subparsec Jets
  in Quasars}, \apj 534 (2000) 109--113.

\bibitem{Celotti2008}
A.~{Celotti}, G.~{Ghisellini}, {The power of blazar jets}, \mnras 385 (2008)
  283--300.

\bibitem{Tavecchio2011}
F.~{Tavecchio}, et~al., {On the origin of the {$\gamma$}-ray emission from the
  flaring blazar PKS 1222+216}, \aap 534 (2011) A86+.

\bibitem{Volonteri2011}
M.~{Volonteri}, et~al., {Blazars in the early Universe}, \mnras 416 (2011)
  216--224.

\bibitem{CGRO-CenA}
H.~{Steinle}, et~al., {COMPTEL observations of Centaurus A at MeV energies in
  the years 1991 to 1995}, \aap 330 (1998) 97--107.

\bibitem{HEGRA-M87}
F.~{Aharonian}, et~al., {Is the giant radio galaxy M 87 a TeV gamma-ray
  emitter?}, \aap 403 (2003) L1--L5.

\bibitem{LAT-MAGN}
A.~A. {Abdo}, et~al., {Fermi Large Area Telescope Observations of Misaligned
  Active Galactic Nuclei}, \apj 720 (2010) 912--922.

\bibitem{VHE-CenA}
F.~{Aharonian}, et~al., {Discovery of Very High Energy {$\gamma$}-Ray Emission
  from Centaurus a with H.E.S.S.}, \apjl 695 (2009) L40--L44.

\bibitem{VHE-IC310}
J.~{Aleksi{\'c}}, et~al., {MAGIC Collaboration}, {Detection of Very High Energy
  {$\gamma$}-ray Emission from the Perseus Cluster Head-Tail Galaxy IC 310 by
  the MAGIC Telescopes}, \apjl 723 (2010) L207--L212.

\bibitem{LAT-CenA-lobes}
A.~A. {Abdo}, et~al., {Fermi-LAT Collaboration}, {Fermi Gamma-Ray Imaging of a
  Radio Galaxy}, Science 328 (2010) 725--.

\bibitem{Takeuchi2012}
Y.~{Takeuchi}, et~al., {Suzaku X-Ray Imaging of the Extended Lobe in the Giant
  Radio Galaxy NGC 6251 Associated with the Fermi-LAT Source 2FGL
  J1629.4+8236}, \apj 749 (2012) 66.

\bibitem{Neronov2007}
A.~{Neronov}, F.~A. {Aharonian}, {Production of TeV Gamma Radiation in the
  Vicinity of the Supermassive Black Hole in the Giant Radio Galaxy M87}, \apj
  671 (2007) 85--96.

\bibitem{Stawarz03}
{\L}.~{Stawarz}, M.~{Sikora}, M.~{Ostrowski}, {High-Energy Gamma Rays from FR I
  Jets}, \apj 597 (2003) 186--201.

\bibitem{Hardcastle-CenA}
M.~J. {Hardcastle}, et~al., {High-energy particle acceleration and production
  of ultra-high-energy cosmic rays in the giant lobes of Centaurus A}, \mnras
  393 (2009) 1041--1053.

\bibitem{Inoue2011}
Y.~{Inoue}, {Contribution of Gamma-Ray-loud Radio Galaxies' Core Emissions to
  the Cosmic MeV and GeV Gamma-Ray Background Radiation}, \apj 733 (2011)
  66--+.

\bibitem{Antonucci2011}
R.~{Antonucci}, {Thermal and Nonthermal Radio Galaxies}, ArXiv e-prints:\href
  {http://arxiv.org/abs/1101.0837} {\path{arXiv:1101.0837}}.

\bibitem{LAT-CenA-core}
A.~A. {Abdo}, et~al., {Fermi Large Area Telescope View of the Core of the Radio
  Galaxy Centaurus A}, \apj 719 (2010) 1433--1444.

\bibitem{kataoka-BLRGs}
J.~{Kataoka}, et~al., {Broad-line Radio Galaxies Observed with Fermi-LAT: The
  Origin of the GeV {$\gamma$}-Ray Emission}, \apj 740 (2011) 29.

\bibitem{Kataoka2007}
J.~{Kataoka}, et~al., {Probing the Disk-Jet Connection of the Radio Galaxy 3C
  120 Observed with Suzaku}, \pasj 59 (2007) 279--297.

\bibitem{Marscher2002}
A.~P. {Marscher}, et~al., {Observational evidence for the accretion-disk origin
  for a radio jet in an active galaxy}, \nat 417 (2002) 625--627.

\bibitem{Sarazin1986}
C.~L. {Sarazin}, {X-ray emission from clusters of galaxies}, Reviews of Modern
  Physics 58 (1986) 1--115.

\bibitem{Blasi2007}
P.~{Blasi}, S.~{Gabici}, G.~{Brunetti}, {Gamma Rays from Clusters of Galaxies},
  International Journal of Modern Physics A 22 (2007) 681--706.

\bibitem{Jeltema2009}
T.~E. {Jeltema}, J.~{Kehayias}, S.~{Profumo}, {Gamma rays from clusters and
  groups of galaxies: Cosmic rays versus dark matter}, \prd 80~(2) (2009)
  023005--+.

\bibitem{Ferrari2008}
C.~{Ferrari}, et~al., {Observations of Extended Radio Emission in Clusters},
  \ssr 134 (2008) 93--118.

\bibitem{Dennison1980}
B.~{Dennison}, {Formation of radio halos in clusters of galaxies from
  cosmic-ray protons}, \apjl 239 (1980) L93--L96.

\bibitem{Magic-clusters}
J.~{Aleksi{\'c}}, et~al., {MAGIC Collaboration}, {MAGIC Gamma-ray Telescope
  Observation of the Perseus Cluster of Galaxies: Implications for Cosmic Rays,
  Dark Matter, and NGC 1275}, \apj 710 (2010) 634--647.

\bibitem{Petrosian2001}
V.~{Petrosian}, {On the Nonthermal Emission and Acceleration of Electrons in
  Coma and Other Clusters of Galaxies}, \apj 557 (2001) 560--572.

\bibitem{Wik2011}
D.~R. {Wik}, et~al., {The Lack of Diffuse, Non-thermal Hard X-ray Emission in
  the Coma Cluster: The Swift Burst Alert Telescope's Eye View}, \apj 727
  (2011) 119--+.

\bibitem{LAT-clusters}
M.~{Ackermann}, et~al., {GeV Gamma-ray Flux Upper Limits from Clusters of
  Galaxies}, \apjl 717 (2010) L71--L78.

\bibitem{HESS-Coma}
F.~{Aharonian}, et~al., {Constraints on the multi-TeV particle population in
  the Coma galaxy cluster with HESS observations}, \aap 502 (2009) 437--443.

\bibitem{Brunetti2011}
G.~{Brunetti}, A.~{Lazarian}, {Acceleration of primary and secondary particles
  in galaxy clusters by compressible MHD turbulence: from radio haloes to
  gamma-rays}, \mnras 410 (2011) 127--142.

\bibitem{Colafrancesco2011}
S.~{Colafrancesco}, et~al., {Dark matter interpretation of the origin of
  non-thermal phenomena in galaxy clusters}, \aap 527 (2011) A80+.

\bibitem{Pinzke2010}
A.~{Pinzke}, C.~{Pfrommer}, {Simulating the {$\gamma$}-ray emission from galaxy
  clusters: a universal cosmic ray spectrum and spatial distribution}, \mnras
  409 (2010) 449--480.

\bibitem{Pfrommer2008}
C.~{Pfrommer}, {Simulating cosmic rays in clusters of galaxies - III.
  Non-thermal scaling relations and comparison to observations}, \mnras 385
  (2008) 1242--1256.

\bibitem{Rephaeli1979}
Y.~{Rephaeli}, {Relativistic electrons in the intracluster space of clusters of
  galaxies - The hard X-ray spectra and heating of the gas}, \apj 227 (1979)
  364--369.

\bibitem{Rephaeli2008}
Y.~{Rephaeli}, et~al., {Nonthermal Phenomena in Clusters of Galaxies}, \ssr 134
  (2008) 71--92.

\bibitem{Ajello2009}
M.~{Ajello}, et~al., {Galaxy Clusters in the Swift/Burst Alert Telescope Era:
  Hard X-rays in the Intracluster Medium}, \apj 690 (2009) 367--388.

\bibitem{Reimer2003}
O.~{Reimer}, et~al., {EGRET Upper Limits on the High-Energy Gamma-Ray Emission
  of Galaxy Clusters}, \apj 588 (2003) 155--164.

\bibitem{Pinzke2011}
A.~{Pinzke}, C.~{Pfrommer}, L.~{Bergstr{\"o}m}, {Prospects of detecting
  gamma-ray emission from galaxy clusters: Cosmic rays and dark matter
  annihilations}, \prd 84~(12) (2011) 123509.

\bibitem{Bykov2000}
A.~M. {Bykov}, H.~{Bloemen}, Y.~A. {Uvarov}, {Nonthermal emission from clusters
  of galaxies}, \aap 362 (2000) 886--894.

\bibitem{Miniati2001}
F.~{Miniati}, et~al., {Cosmic-Ray Protons Accelerated at Cosmological Shocks
  and Their Impact on Groups and Clusters of Galaxies}, \apj 559 (2001) 59--69.

\bibitem{Timokhin2004}
A.~N. {Timokhin}, F.~A. {Aharonian}, A.~Y. {Neronov}, {On the non-thermal high
  energy radiation of galaxy clusters}, \aap 417 (2004) 391--399.

\bibitem{Fujita2007}
Y.~{Fujita}, K.~{Kohri}, R.~{Yamazaki}, M.~{Kino}, {Nonthermal Emission
  Associated with Strong AGN Outbursts at the Centers of Galaxy Clusters},
  \apjl 663 (2007) L61--L64.

\end{thebibliography}

\end{document}